\documentclass[epj]{svjour}
\usepackage{amsmath,amssymb}
\usepackage[export]{adjustbox}
\usepackage{color}
\usepackage{graphicx}% Include figure files
\usepackage{dcolumn}% Align table columns on decimal point
\usepackage{bm}% bold math
\usepackage{hyperref}% add hypertext capabilities
\usepackage{float}
\usepackage[caption=false]{subfig}
\usepackage{subfloat}
\usepackage{hyperref}
\hypersetup{
colorlinks=true,
linktoc=page,    %set to all if you want both sections and subsections linked
linkcolor=blue,
citecolor=blue,
urlcolor=blue,%magenta,
colorlinks=true,
}

\newcommand{\be}{\begin{equation}}
\newcommand{\ee}{\end{equation}}
\newcommand{\bea}{\begin{eqnarray}}
\newcommand{\eea}{\end{eqnarray}}

\newcommand{\ZZ}{\mathbb{Z}}

\newcommand{\mD}{\mathcal{D}}

\newcommand{\tr}{\text{tr}}

\def\/{\frac}

\newcommand{\dd}{\,\mathrm{d}}

\newcommand{\bra}[1]{\left<{#1}\right|}
\newcommand{\brat}[2]{\left<{#1}|{#2}\right>}
\newcommand{\ketbra}[1]{|{#1}\left>\right<{#1}|}
\newcommand{\ket}[1]{\left|{#1}\right>}
\newcommand{\braket}[2]{\left<{#1}\right|{#2}\left|{#1}\right> }
\def\bracket#1{\left\langle #1 \right\rangle}

\begin{document}

%Local setting
\title{Notes on Wormhole Cancellation and Factorization}
\author{Peng Cheng\inst{a,b}, Pujian Mao\inst{a}}
\institute{${}^a$Center for Joint Quantum Studies and Department of Physics,
     School of Science, Tianjin University, 135 Yaguan Road, Tianjin 300350, China\\
     ${}^b$Lanzhou Center for Theoretical Physics, Key Laboratory of Theoretical Physics of Gansu Province, and Key Laboratory of Quantum Theory and Applications of MoE, Lanzhou University, Lanzhou, Gansu 730000, China
}
\date{}

%\email{p.cheng.nl@outlook.com, pjmao@tju.edu.cn}   
\abstract{
In AdS/CFT, partition functions of decoupled CFTs living on separate asymptotic boundaries factorize. 
However, the presence of bulk wormholes connecting different boundaries tends to spoil the factorization of the bulk partition function, which leads to a disagreement between the two sides.
In this paper, we present two examples where wormhole contributions cancel each other in bulk partition function calculations, thus the bulk factorization can be realized. 
The first example is in 2-dimensional Jackiw-Teitelboim (JT) gravity, where the proposed way of realizing the cancellation resides in the extra complex phases associated with different wormholes. The phases arise due to the degenerate vacua structure.
In the example of the Sachdev-Ye-Kitaev (SYK) model, the cancellation can be achieved due to the distribution of the wormhole saddles on a complex plane.
The two examples demonstrate a way of realizing bulk partition function factorization by extending the Hilbert space and dressing wormhole saddles with extra phases.
}

\maketitle

%%%%%%%%%%%%%%%%%%%%%%%%%%%%%%%%%%%%%%%%%%%%%%%%%%%%%%%%%%%%%%%%%%%%%%%%%%%%%%%%%%%%%%%%%%%%%%%%%%%%

%\tableofcontents
%\flushbottom
%\newpage

%%%%%%%%%%%%%%%%%%%%%%%%%%%%%%%%%%%%%%%%%%%%%%%%%%%%%%%%%%%%%%%%%%%%%%%%%%%%%%%%%%%%%%%%%%%%%%%%%%%%
% MAIN BODY
%%%%%%%%%%%%%%%%%%%%%%%%%%%%%%%%%%%%%%%%%%%%%%%%%%%%%%%%%%%%%%%%%%%%%%%%%%%%%%%%%%%%%%%%%%%%%%%%%%%%
\section{Introduction}
\label{intro}
%\input{sec/1-intro.tex}

%\section{Introduction}
%\label{intro}

%%%%%% the factorization puzzle %%%%%%
The recent progress on low-dimensional holography \cite{Saad2019,AfkhamiJeddi2020,Maloney2020} and the black hole information paradox \cite{Penington2019a,Almheiri2019b} have brought challenges to the standard AdS/CFT correspondence.
As an example, in the derivation of the so-called island formula \cite{Penington2019,Almheiri2019,Almheiri2019a}, wormhole geometries connecting different asymptotic boundaries and higher genus geometries were introduced in the Euclidean path integral \cite{Penington2019a,Almheiri2019b,Kundu:2021nwp}.
The inclusion of bulk connecting geometries causes an apparent puzzle: the factorization puzzle\footnote{Note that there is a slightly different but related factorisation puzzle.
According to the Maxfield-Penington-Witten convention \cite{Penington:2023dql}, the spelling ``factorization'' specifically means the puzzle that spacetime wormholes cause partition functions not to factorize on a set of disconnected asymptotic spacetime boundaries.
While the factorisation puzzle is a Lorentzian description where concepts like Hilbert space and algebras are involved.
} \cite{Maldacena2004,Harlow:2018tqv}.

The factorization puzzle is a tension between the bulk gravitational partition function and dual quantum theory in the AdS/CFT.  
Boundary theories living on several disconnected asymptotic regions naturally produce factorized partition functions, this is because actual boundary partition function calculation usually does not include non-local effects.
However, the boundary factorization is in conflict with the existence of bulk wormholes. 
In the presence of bulk wormholes connecting different boundaries, the bulk partition function obviously doesn't factorize.
The puzzle is essentially the disagreement between the boundary and bulk perspectives.

%%%%%%% The puzzle %%%%%%

%Due to the presence of bulk wormholes or other geometries that connect different boundaries, there is a factorization puzzle in the theory\footnote{We are going to use term ``wormhole'' to denote all the bulk geometries connecting different boundaries.}. 
%The puzzle is a tension between the boundary quantum theory and the bulk gravity partition function in holography.
To see the puzzle more explicitly, assuming we have $n$ non-connected $d$-dimensional boundary components labelled by $\{B_1,B_2,\cdots,B_i,\cdots,B_n\}$. 
Since the $n$ parts of spacetime are non-connected, the locality of the boundary field theory implies that the overall partition function should factorize to the products of $n$ partition functions, which can be formally denoted as 
\be
Z_{\text{bdy}}=\prod_{i=1}^n Z[B_i]\,.
\ee 
Note that we have denoted the partition function on each component $B_i$ as $Z[B_i]$.
Now, if each $d$-dimensional component is a boundary dual of a bulk gravity theory, according to the  Gubser-Klebanov-Polyakov-Witten (GKPW) relation \cite{Gubser:1998bc,Witten:1998qj}, the bulk partition function equals the boundary partition function, i.e. $Z_i^{\text{g}}=Z[B_i]$. Partition functions can be derived using Euclidean path integral on the Euclidean manifold, so we can use the corresponding Euclidean geometry to illustrate the bulk partition function, as
\bea\label{holo}
Z[B]~~\overset{\text{holo}}{=}~~
\begin{matrix}
\includegraphics[width=1.3cm]{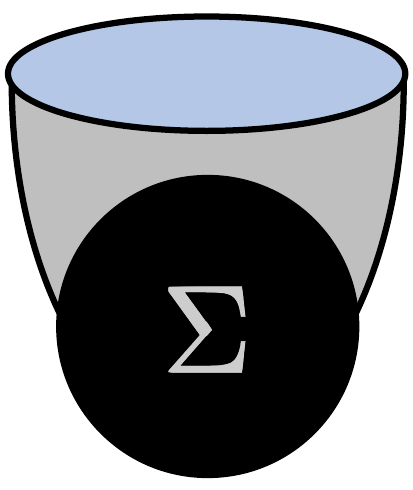}
\end{matrix}
\eea
where we have used $\Sigma$ to represent the summation of high-genus geometries\footnote{We have used $\Sigma$ to represent the following diagrams
\begin{equation*}
\begin{matrix}
\includegraphics[width=0.9cm]{pic/sig0}
\end{matrix}
=\begin{matrix}
\includegraphics[width=0.9cm]{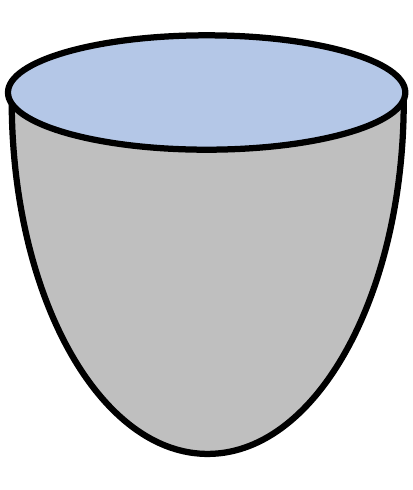}
\end{matrix}
+
\begin{matrix}
\includegraphics[width=0.8cm]{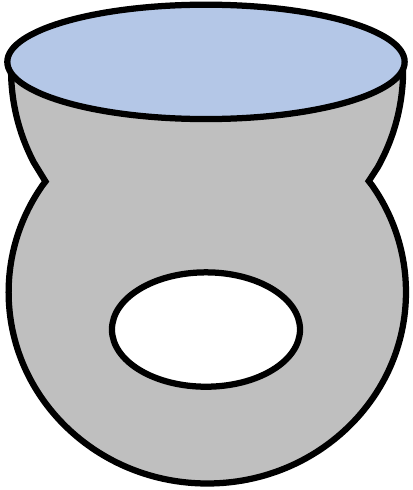}
\end{matrix}
+
\begin{matrix}
\includegraphics[width=0.8cm]{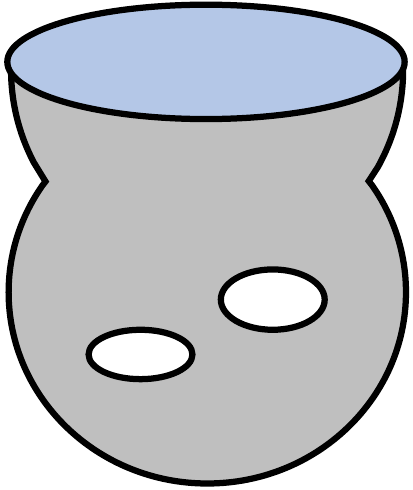}
\end{matrix}+\cdots
\end{equation*}
}.
Due to the traditional holographic dictionary, it is natural to conclude that the bulk gravitational partition function should also be factorized into $n$ pieces
\be
Z_{\text{gravity}}=\prod_{i=1}^nZ_i^{\text{g}}=\prod_{i=1}^n Z[B_i]=Z_{\text{bdy}}\,,
\ee
which can be illustrated as
%\bea
%Z_{\text{gravity}}~~
%=~~
%\begin{matrix}
%\includegraphics[height=2.5cm]{pic/sigma1}
%\end{matrix}\,.
%\eea
\bea
Z_{\text{gravity}}~~
=~~
\begin{matrix}
\includegraphics[height=2.6cm]{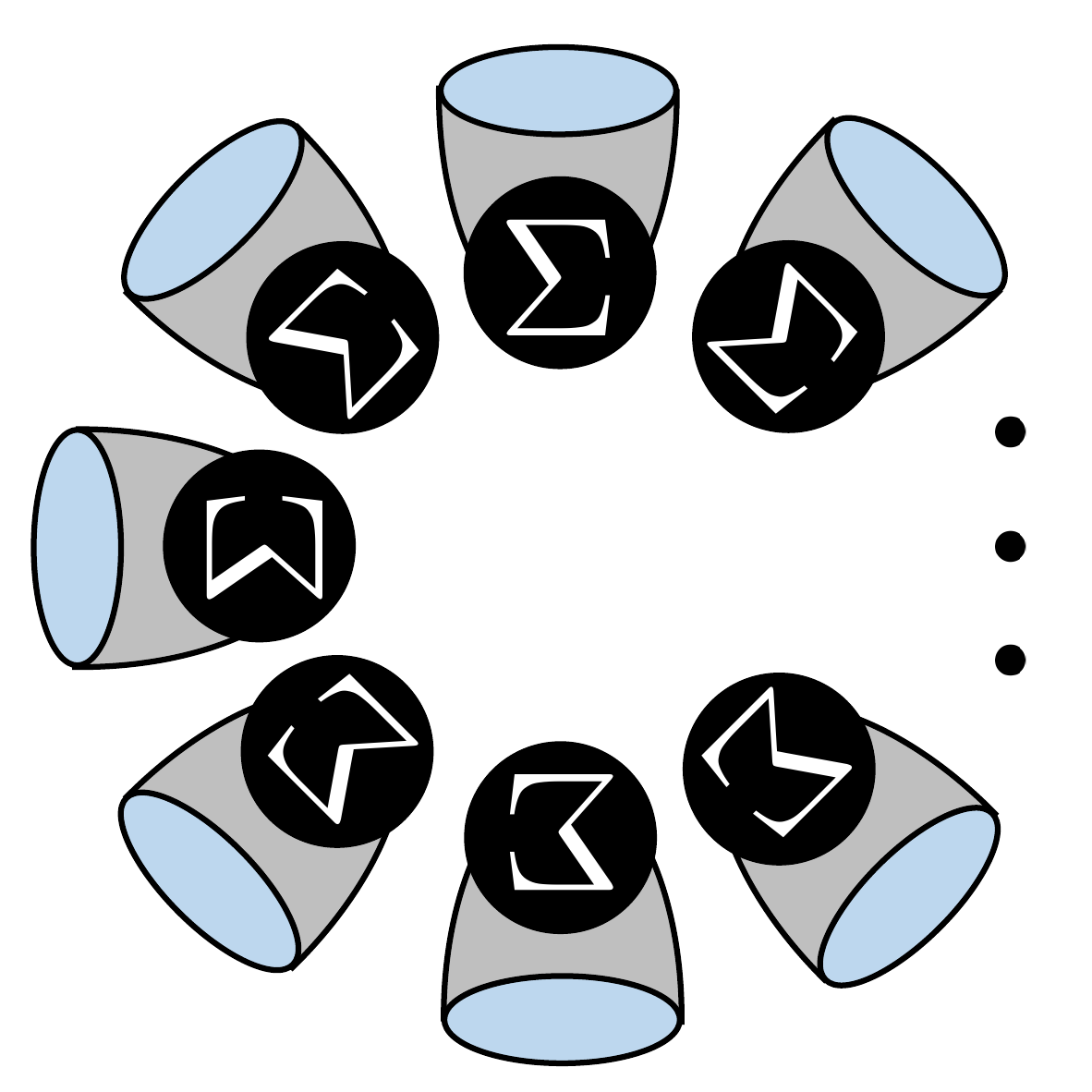}
\end{matrix}\,.
\eea
However, in the presence of bulk wormholes connecting different boundaries, like the geometries shown below
\bea
\begin{matrix}
\includegraphics[height=2.6cm]{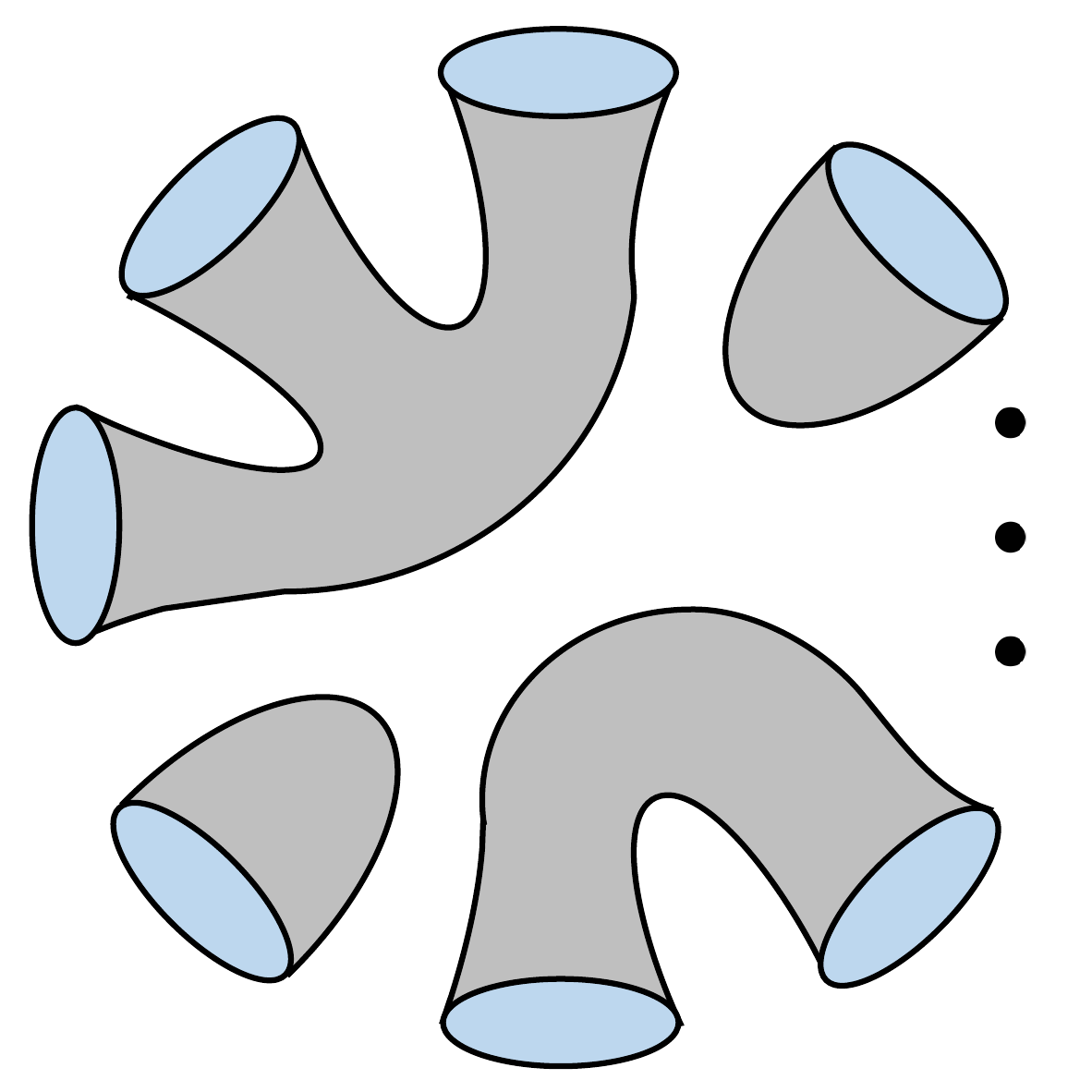}
\end{matrix}\,,
\eea
the bulk partition function naively does not factorize and thus does not equal the boundary partition function. 
So, we have seen a problem due to the presence of the bulk connecting geometries.

Note that we have assumed that non-local effects are not included in boundary partition function calculations.
Starting with decoupled boundary components, while one finds wormholes in bulk geometry. This is the origin of the factorization puzzle. 
One can also include non-local interactions in the boundary theories, such that factorization is not a necessary property. 
Non-local interactions may cause further confusion when doing path integral that we do not intend to deal with.
We mainly focus on boundary theories without non-local effects and try to find bulk theories that respect the factorization property in this paper.

%Assuming we have $n$ non-connected $d$-dimensional boundary components labelled by $\{B_1,B_2,\cdots,B_n\}$, each component $B_i$ are supposed to have a partition function denoted as $Z[B_i]$. So, the boundary partition function can be formally denoted as $\prod_{i=1}^n Z[B_i]$. Now let us consider a $d+1$ dimensional bulk gravity theory with $n$ boundaries $\{B_1,B_2,\cdots,B_n\}$. The bulk partition function doesn't factorize because of the possible wormholes and would never equal the boundary partition function $\prod_{i=1}^n Z[B_i]$. 

%%%%%% Attempts %%%%%%
There are several excellent attempts to understand the factorization puzzle.
It was suggested that the gravity theory might be dual to an ensemble of theories \cite{Saad2019,AfkhamiJeddi2020,Maloney2020,Penington2019a,Marolf2020a,Coleman1988,Giddings1988,Giddings1989,Polchinski1994,Maldacena2016a,Jensen2016,Heckman:2021vzx}, where factorization is not a necessary property because of the possible variance.
Dual to ensemble-averaged boundary theories is different than the standard AdS/CFT correspondence where we only have a specific boundary theory. So the above proposal raises alternative confusion about which one is the right duality, the ensemble-averaged one or the standard one. 
Recent studies suggest that factorization can be restored if the half-wormhole contributions are added in \cite{Saad2021a,Saad2021,Peng2021,Peng:2022pfa,Blommaert2021}.
However, the factorization requires a nontrivial relationship between the contributions of wormholes and half-wormholes, where some strange bi-local interactions between the half-wormholes are also added into the story \cite{Blommaert2021}.
%All the above facts make the proposal less natural, and the half-wormholes and bi-local interactions still need further understanding.
In \cite{Benini2022}, Benini \textit{et al.} show that by gauging a global 1-form symmetry in 3-dimensional bulk Chern-Simons theory, the dimension of the bulk baby universe Hilbert space is one and the factorization can be realized. 
The relation between symmetry breaking and bulk topology is also hinted at there.
%Although factorization can be restored, we are not sure whether completely killing the bulk topological modes is what we wanted or not. Hence, a convincing resolution for the factorization puzzle is still on its way and it is the main motivation of the present work.
All the above results suggest that the factorization puzzle can be restored when the bulk theory is equipped with some special structure. 
The aim of the present work is to further explore what structure can be used to understand the factorization puzzle.
%which can lead to a factorized bulk theory.
%One of the crucial differences between the ensemble-averaged and the standard correspondence is the necessity of the factorization property. A good comprehension of the factorization puzzle seems to be vital to the holographic duality.

The factorization puzzle is a problem due to the wormhole geometries connecting different boundaries. 
But it is especially conspicuous and high-profile in topological gravity theories, such as the 2-dimensional Jackiw-Teitelboim (JT) gravity \cite{Jackiw1985,Teitelboim1983,Teitelboim1983a} and the related Sachdev-Ye-Kitaev (SYK) model \cite{Sachdev1993,Kitaev2015,Kitaev2017}. In high dimensions, the bulk high-genus geometries are largely suppressed due to the exponential suppression in genus expansion. 
In this paper, we are going to show two examples where the wormhole contributions cancel each other and the bulk factorization can be realized. The examples are in the JT gravity and the SYK model.  
In the low-dimensional examples, there is no contrast between the extra parameter and the McNamara-Vafa baby universe hypothesis \cite{McNamara:2020uza}.

In JT gravity, when degenerate vacua is included in the theory, the bulk geometries connecting different boundaries are associated with extra phases. Those phases can ensure the cancellation of those geometries and we are left with disconnected geometries. The cancellation provides a way of realizing bulk partition function factorization.
In the SYK model, the wormhole saddles are associated with a complex phase, and their distribution on the complex plane leads to a similar wormhole cancellation. Connections between the two examples remain unclear and need further studies.
When a specific set of boundary vacua is chosen in the JT gravity (or one single wormhole saddle is designated in the SYK model), all the bulk connecting geometries (or interactions between different replicas) can be seen, which can be regarded as the ``classical limit''.
This explains what ingredient is missing in the theories with a factorization puzzle. The degenerate vacua structure might be the necessary ingredient for bulk partition function factorization.

%%%%%% The structure of this paper %%%%%%
The paper is organized as follows. 
%\textcolor{red}{The section \ref{fac} is a review of the factorization puzzle.}
In section \ref{bulk}, we study the cancellation between the bulk geometries connecting different boundaries when degenerate vacua is considered.
Section \ref{dual} mainly focuses on the wormhole saddle points in the one-time point SYK model, and demonstrates a similar cancellation in the SYK model. 
Section \ref{conc} is the conclusion and discussion section. 
In appendix \ref{AA}, we reviewed the Euclidean path integral and the origin of the degenerate vacua. Appendix \ref{BB} illustrates the concept of wormholes in the SYK model with fixed coupling.

%%%%%%%%%%%%%%%%%%%%%%%%%%%%%%%%%%%%%%%%%%%%%%%%%%%%%%%%%%%%%%%%%%%%%%%%%%%%%%%%%%%%%%%%%%%%%%%%%%%%

%%\input{sec/2-puzzle.tex}
%\section{The factorization puzzle}
%\label{fac}

%%%%%%%%%%%%%%%%%%%%%%%%%%%%%%%%%%%%%%%%%%%%%%%%%%%%%%%%%%%%%%%%%%%%%%%%%%%%%%%%%%%%%%%%%%%%%%%%%%%%

%\newpage
%\input{sec/3-bulk}
\section{Wormhole cancellation in JT gravity with extra vacua}
\label{bulk}

%%%%%%% A brief intro of the JT gravity %%%%%%

We will demonstrate the wormhole cancellation in the 2-dimensional JT gravity in this section.
The JT gravity theory consists of a 2-dimensional metric $g_{\mu\nu}$ and a dilaton field $\Phi$, whose action can be written as
\be\label{JT}
I[g,\Phi]=-S_0 \chi -\frac{1}{2}\int_\mathcal{M}\sqrt{g}~\Phi(R-2\Lambda)-\int_{\partial \mathcal{M}}\sqrt{h}~\Phi(K-1)\,,
\ee
where $\chi$ is the Euler characteristic. 
$S_0$ can be regarded as the constant mode of $\Phi$, or extremal black hole entropy if the theory is reduced from high dimensional near horizon geometry. 
The variation of the dilaton field $\Phi$ results in $R-2\Lambda=0$. 
Then, the bulk action is only left with the topological part and doesn't have any bulk propagating degrees of freedom. 
The remaining degrees of freedom are all on the boundary, where we have $\Phi|_{\partial \mathcal{M}}=C/\epsilon$ and $g_{uu}|_{\partial \mathcal{M}}=1/\epsilon^2$ as the boundary condition. 
It can be shown that the boundary theory is described by the Schwarzian action \cite{Maldacena2016}
\be\label{Schwarzian}
S[f]=-\frac{1}{2}\int dt~ \{f,t\}\,,~~~\text{with}~ \{f,t\}= \frac{f'''}{f''}-\frac{3}{2}\left(\frac{f''}{f'}\right)^2\,.
\ee
Now, we have a topological gravity theory, whose boundary theory is described by the Schwarzian theory \eqref{Schwarzian}. 
The Schwarzian theory, as well as the 2-dimensional gravity theory, can be regarded as the low energy limit of the SYK model, which is a very useful tool to clarify lots of conceptual problems.
%Thus, the JT gravity and the SYK model together provide ideal toy models to understand the low-dimensional system.

In 2-dimensional gravity theories, the boundaries are 1-dimensional circles. We can use $Z[ \bigcirc]$ to denote the partition function on a boundary circle. For simplification, if we consider the case with two asymptotic boundaries $\{B_1, B_2\}$, the factorization puzzle can be summarized as the inequality between the boundary factorized partition function and the partition function of the bulk geometries, which can be illustrated as follows
\bea\label{puzzle}
Z[ \bigcirc]\times Z[\bigcirc ]~~
\neq~~
\begin{matrix}
\includegraphics[height=2.5cm]{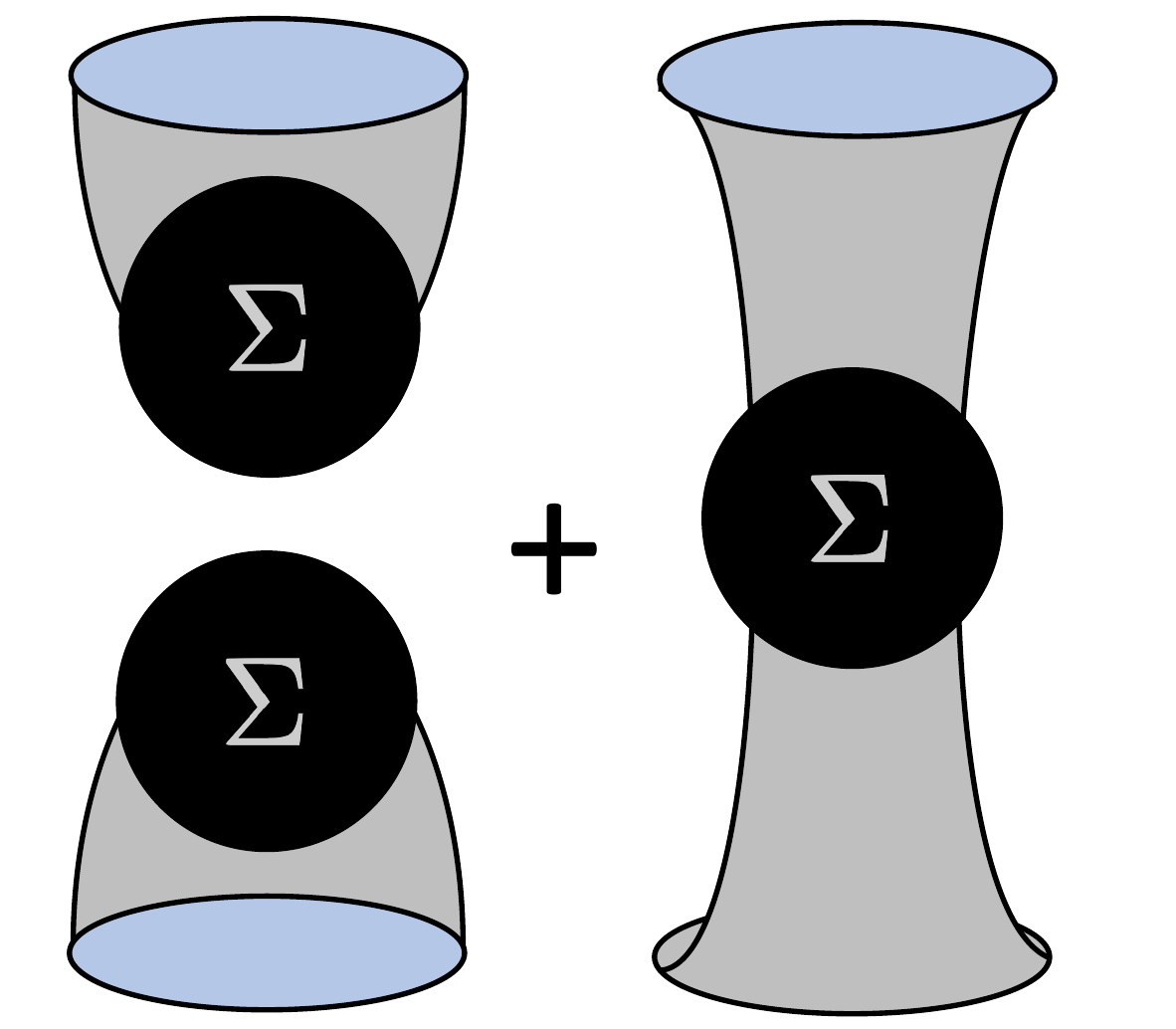}
\end{matrix}\,.
\eea
In order to denote the puzzle shown in \eqref{puzzle} in text, we are going to use $Z\left[\bigcirc \cup \bigcirc \right]$ to denote the right part of \eqref{puzzle} when $n=2$, and express the puzzle as inequality
\bea
Z[ \bigcirc]\times Z[\bigcirc ]~
\neq~
Z\left[\bigcirc \cup \bigcirc \right]\,.
\eea

From Eq. \eqref{puzzle}, the solution to the puzzle seems pretty obvious. If all the connected geometries cancel each other and give zero in the partition function, then the puzzle can be avoided.
Claiming all connected geometries are subleading seems not to be the right way out. We need to add extra ingredients to the wormhole-like geometries.
But what structure can ensure the cancellation between the wormholes? We are going to show how the cancellation can be realized with the help of degenerate vacua.

%Firstly, we will show the transition between boundaries in different vacua in the low dimensional theory. Then, we will demonstrate the cancellation of bulk geometries connecting different boundaries, starting from the simple cases and then turning to the most general case after that. 

\subsection{Extra vacua}
\label{bulk-1}

It was demonstrated in \cite{An:2023dmo} that the asymptotic boundaries of the wormhole can be in different vacua. To show degenerate vacua can be the key ingredient in understanding the factorization puzzle, we first need to understand the origin and the basic structure of the vacua.

From thermal field theory, the partition function of a system can be derived from the Euclidean path integral on the corresponding Euclidean manifold. That is the reason why we always use the Euclidean manifold to denote the corresponding partition function. For example, the geometry of a Euclidean black hole is a cigar, and the black hole thermodynamics can be derived from a path integral on the cigar geometry. 
Recall that in black hole thermodynamics, the black hole partition function can be obtained by tracing two thermo-field double (TFD) states \cite{Hartman2015,Maldacena_2003}, i.e. $Z=\tr(\ket{\text{TFD}}\bra{\text{TFD}})$. Diagrammatically, the cigar geometry can be obtained by gluing two TFDs, as shown in appendix \ref{AA}.

Similar to the black hole case, the wormhole partition function can be obtained by tracing the thermo-mixed double (TMD) state \cite{Verlinde2020,Verlinde2021,Verlinde2021a}, which can be illustrated as
\be
\text{tr}[\rho^2_{\text{TMD}}] =\sum_{\psi_1,\psi_2}%\sum_{\phi,\bar \phi}
\begin{matrix}
\includegraphics[height=2.7cm]{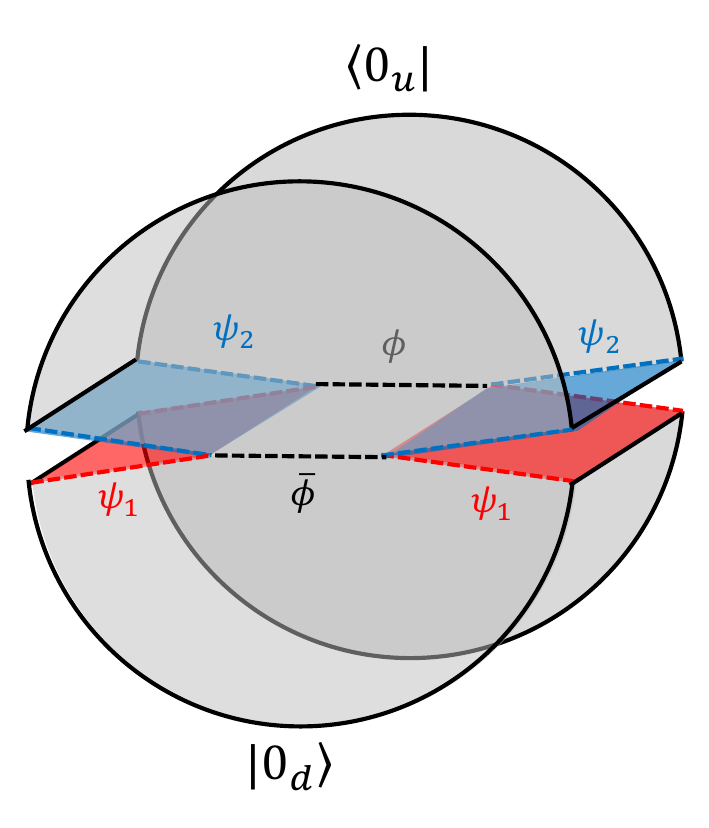}
\end{matrix}
=
\begin{matrix}
\includegraphics[height=2.7cm]{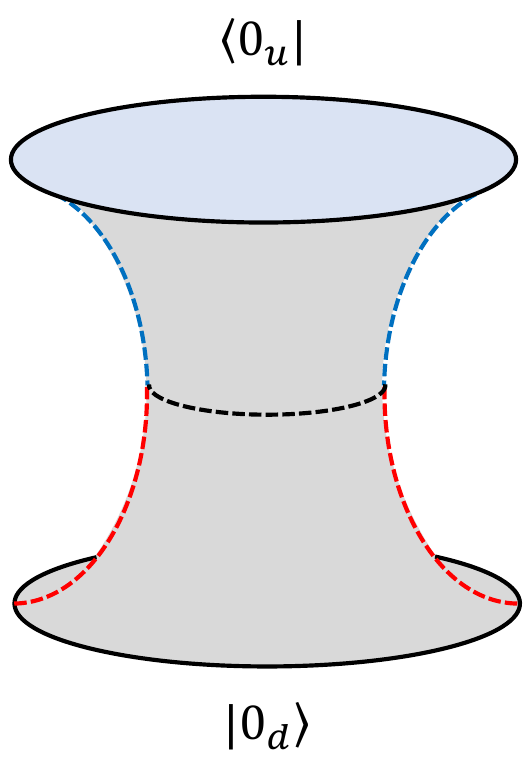}
\end{matrix}\,.\label{vac}\vspace{-0.5em}
\ee
The TMD state can be regarded as a generalization of TFD. 
However, as detailed in appendix \ref{AA}, there are crucial differences between the Euclidean black hole and Euclidean wormhole geometry.
In the cigar case, the trace operation identifies the vacua evolved from $\tau=-\infty$ and $\tau=\infty$ such that there is a unique vacuum for the whole black hole.
However, for the wormholes, there is no trace operation identifying the asymptotic vacua on different boundaries, which means that the boundaries of the wormhole can be in different vacua. 

In 2-dimensional JT gravity, it was shown that the vacuum degeneracy for wormholes can be understood by the twist between the trumpets \cite{An:2023dmo}. Three schematic diagrams are shown below to illustrate the twist and different vacua:
\be\label{3WH}
\begin{matrix}
 \includegraphics[width=2cm]{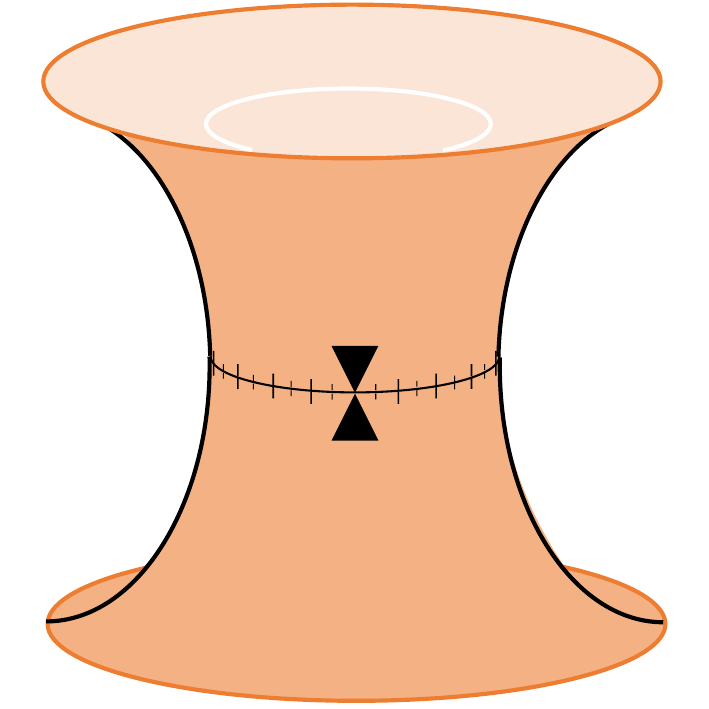}\\(a)
\end{matrix}~~~~~~~~
\begin{matrix}
 \includegraphics[width=2cm]{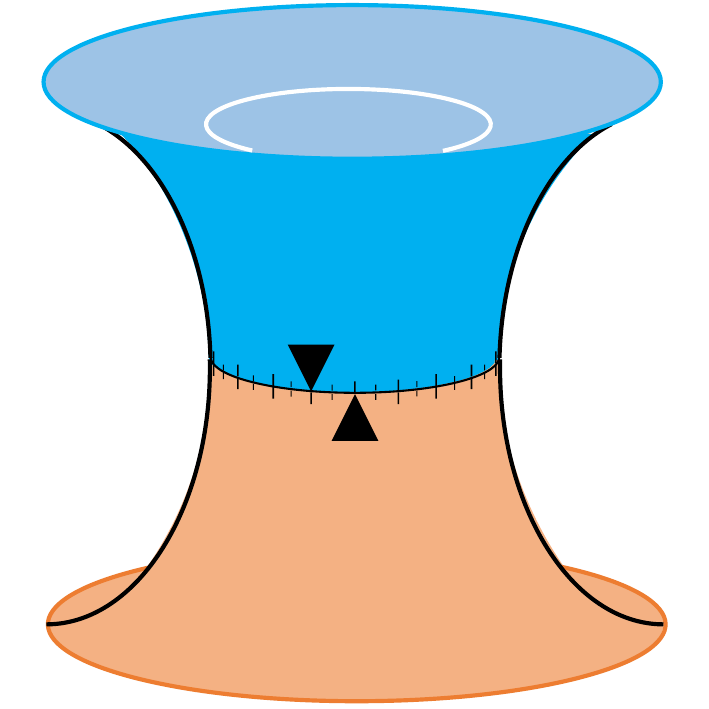}\\(b)
\end{matrix}~~~~~~~~
\begin{matrix}
 \includegraphics[width=2cm]{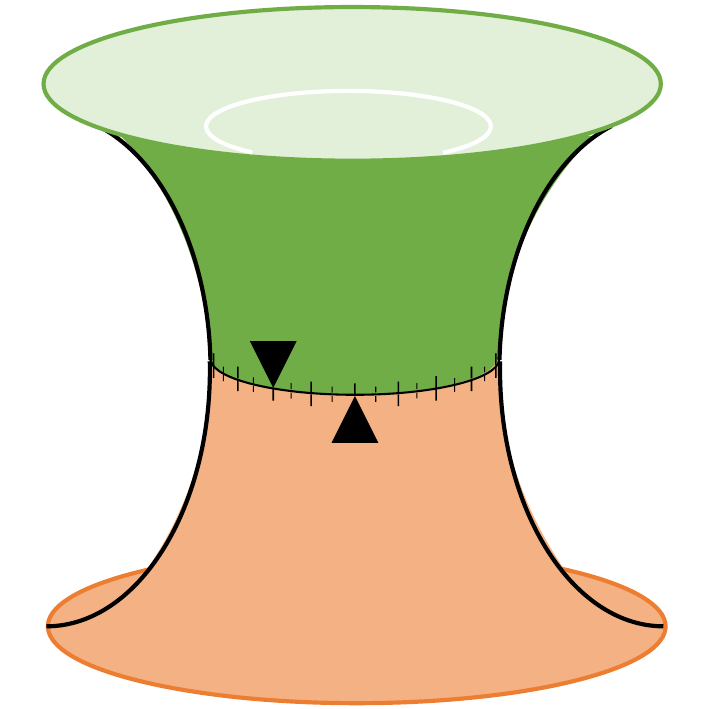}\\(c)
\end{matrix}\,.
\ee
As can be seen from \eqref{3WH}, diagram $(a)$ is the case when we have the same vacuum for both up and down trumpets. Diagrams $(b)$ and $(c)$ are twisted ones, and we can say that the up and down trumpets are in different vacua.

To study the consequence of those degenerate vacua, let us suppose there are $k$ degenerate vacua, labeled by
\be\label{vacua}
\ket{\text{vacua}}=\{\ket{0},\ket{1},\ket{2},\cdots, \ket{k-1}\}\,.
\ee
The vacua are organized according to the cyclic group $\ZZ_k$ with an integer $k$, which means that an operator $\mathcal{S}$ acting on $\ket{i}$ gives out $\ket{i+1}$. $\ket{0}$ can also be obtained by acting the operator on $\ket{k-1}$. %, which means that $\ket{0}=\ket{k}$.
In \eqref{3WH}, the action $\mathcal{S}$ is used to twist $(b)$ back to $(a)$. As we will see later, the cost of twisting $(b)$ back would be an extra factor $S$ related to the twisted angle. So, the extra factor should be regarded as the expectation value of operator $\mathcal{S}$.
Operator $\mathcal{S}$ twisting $(b)$ back to $(a)$ can be formally expressed as
\be
\brat{i}{i+1}=\bra{i}\mathcal{S}\ket{i}=S\times\brat{i}{i}\,.
\ee
In JT gravity, the twist is characterized by a continuous U(1) parameter, while we are going to use a relatively simpler version of U(1) symmetry, i.e. $\ZZ_k$, to demonstrate the mechanism of realizing factorization.
We believe that $\ZZ_k$ symmetry is good enough to capture the essence. Moreover, it can be demonstrated that the mechanism works also in the $k\to \infty$ limit.

Unlike painting the whole diagram in different colors in \eqref{3WH}, we will only use different colors to decorate boundaries in different vacua. Black boundaries are specially reserved to denote the superposition of different vacua when too many diagrams are needed to illustrate all different vacua explicitly.

%We are going to consider the bulk gravity with $n$ asymptotic boundaries each of which can be in any of the $k$ vacua. We will calculate the partition function in the 2-dimensional gravity theory shown in Eq. \eqref{JT}. After the calculation, the key question we are going to answer is when those degenerate vacua are considered, whether the gravity partition function gives out the factorizing result or not?

Like the vacua generated by the soft charge, those different degenerate vacua should not form ``super-selection sector'', and when we compare two different vacua, we would get a number representing the overlap between the vacua.
So, the bulk geometry connecting different vacua equals the geometry connecting boundaries in the same vacuum multiplied by an extra factor $F(S)$.
As the simplest example, the 2-fold Euclidean wormhole connecting $\ket{i}$ and $\ket{i+1}$ can be represented as
\be\label{wh}
\begin{matrix}
\includegraphics[height=3cm]{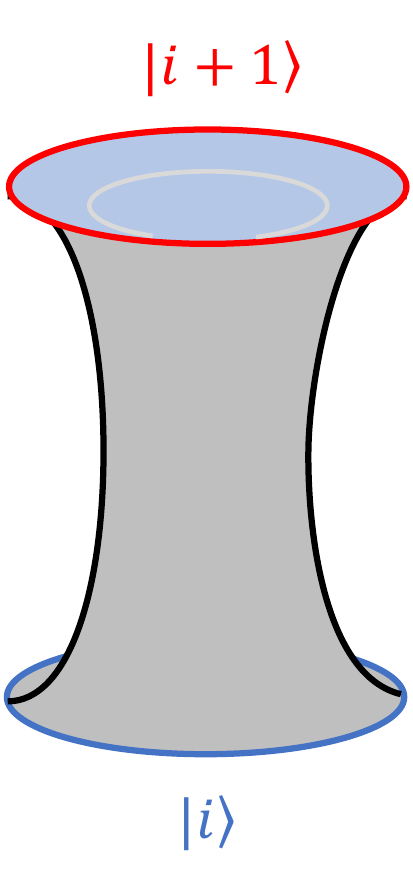}
\end{matrix}~
=~
S~\times ~\begin{matrix}
\includegraphics[height=3cm]{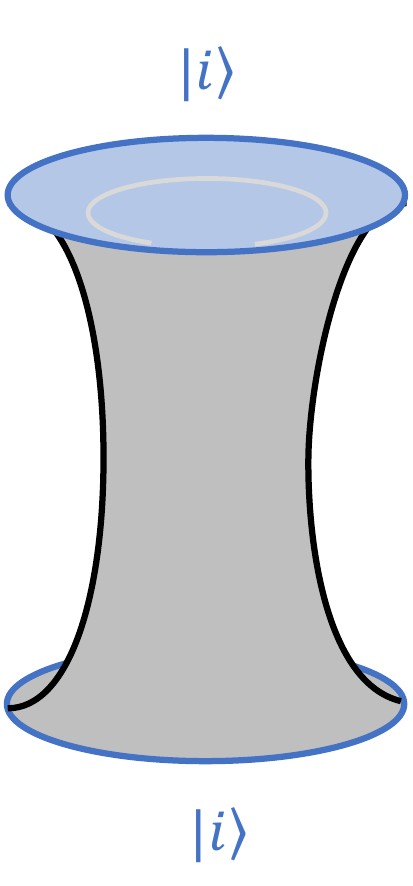}
\end{matrix}
\,
\ee
where $F(S)=S^1$ because $\ket{i+1}$ is only ``one step" away from $\ket{i}$.
%The above prescription makes perfect sense because we can regard the left process in \eqref{wh} as two successive processes: the bulk process between two boundaries both in $\ket{i}$ and then transition from $\ket{i}$ to $\ket{i+1}$.
%In a sense, the extra factor $S$ can be regarded as comparing $\ket{i}$ and $\ket{i+1}$ states
%\be
%S=\braket{i}{\mathcal{S}}\,.
%\ee
%\be
%S=\brat{i+1}{i}\,.
%\ee
The situation when we have nonadjacent vacua $\ket{i+a}$ and $\ket{i}$ shouldn't be hard to obtain, which reads as
\be
\begin{matrix}
\includegraphics[height=3cm]{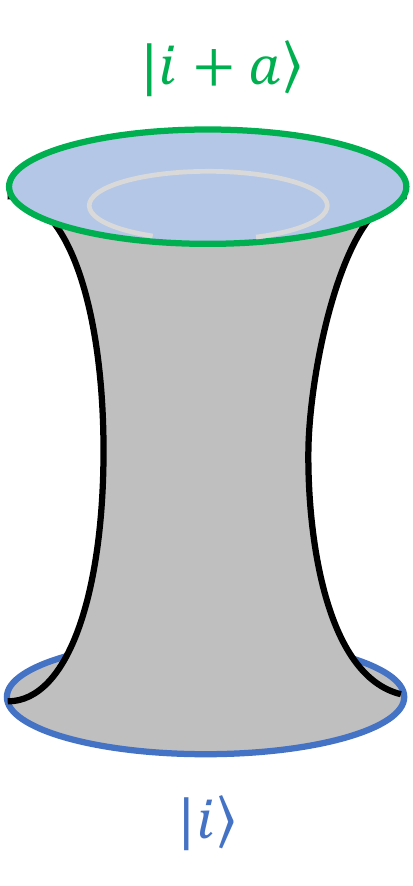}
\end{matrix}~
=~
S^a~\times ~\begin{matrix}
\includegraphics[height=3cm]{pic/wh2}
\end{matrix}\,.
\ee
The above argument can also be generalized to the case where we have more asymptotic boundaries. For example, for the case $n=3$, we have
\be
\begin{matrix}
\includegraphics[height=2.5cm]{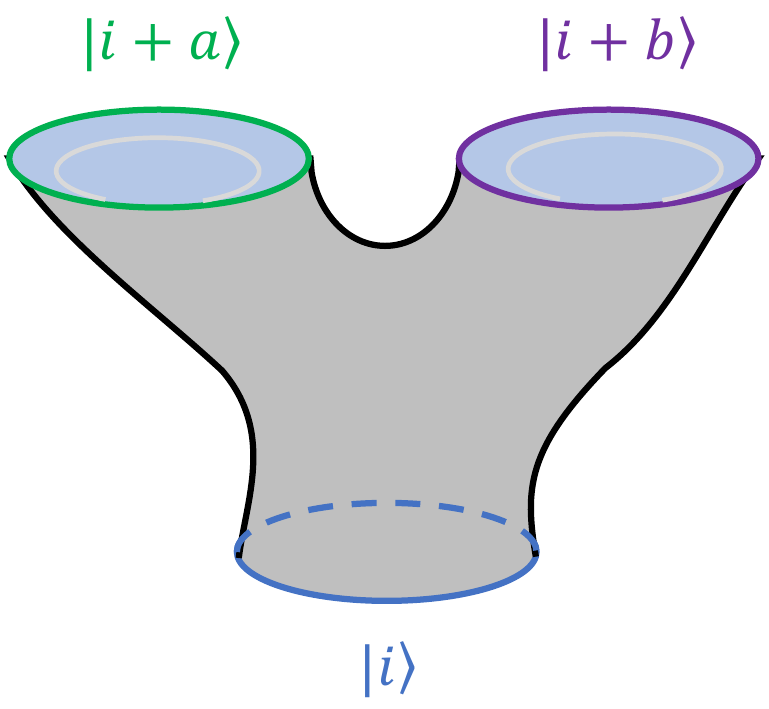}
\end{matrix}~
=~
S^a\times S^b\times ~\begin{matrix}
\includegraphics[height=2.5cm]{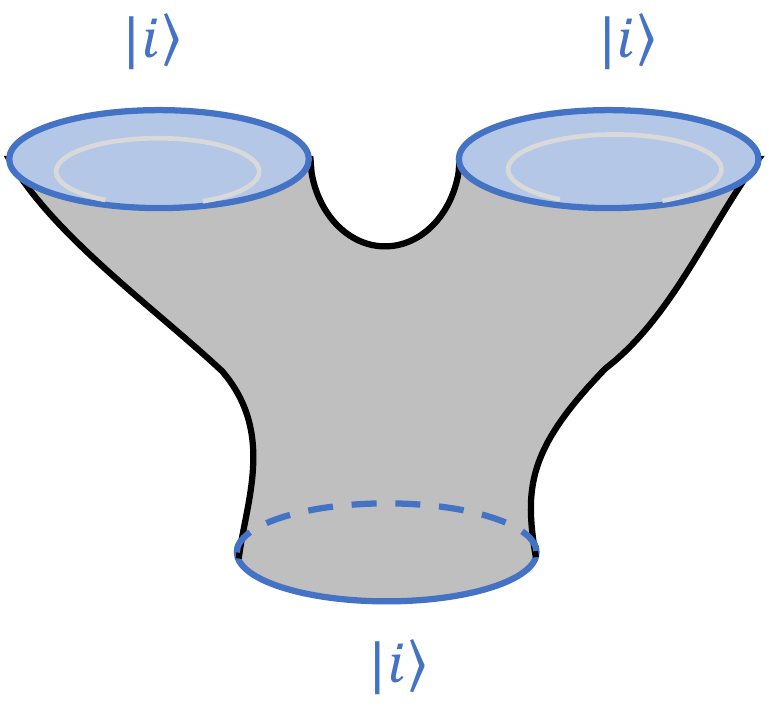}
\end{matrix}\,
\ee
where we get the product of two factors $S^a$ and $S^b$. %The two factors are there because we compare vacuum $\ket{i}$ with $\ket{i+a}$ and $\ket{i+b}$.
Note that the extra factor $S$ only shows up when we are comparing the boundary vacua in one connected diagram. 
There is not a factorized result for disconnected geometries and one needs to keep all the boundary vacua unchanged, because we can not use the overlap between the vacua to further rewrite them.
The above fact would finally cancel the connected geometries and result in the factorized partition function.

Let's use the simplest case ($n=2$, $k=2$) to explain the above expressions more explicitly.
With the degenerate vacua being considered, we get four copies of all bulk geometries with all possible topologies
\bea\label{sum}
Z\left[ \bigcirc\cup  \bigcirc \right]
&=& \begin{matrix}
\includegraphics[height=2.7cm]{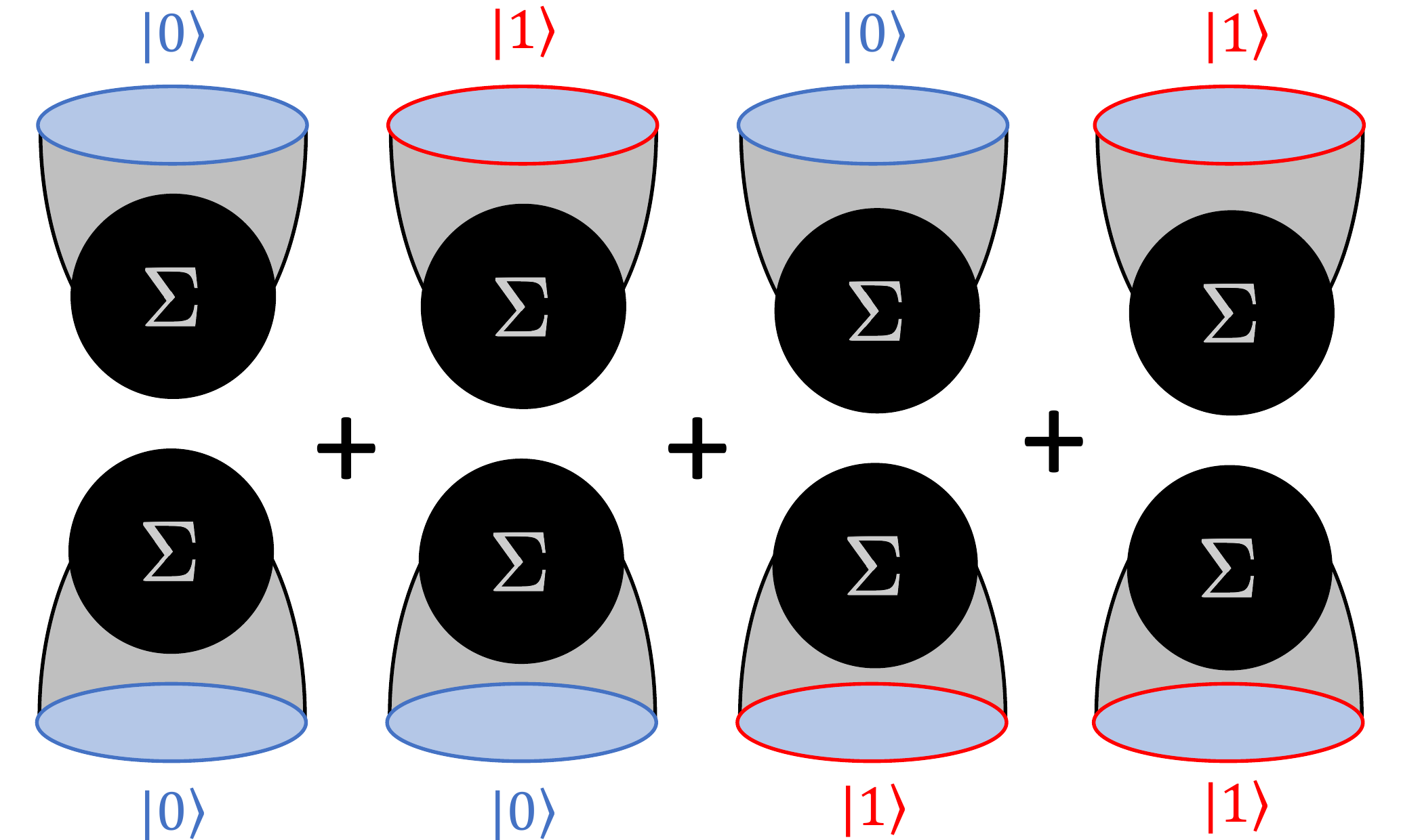}
\end{matrix}\nonumber \\&~&
+ \begin{matrix}
\includegraphics[height=2.7cm]{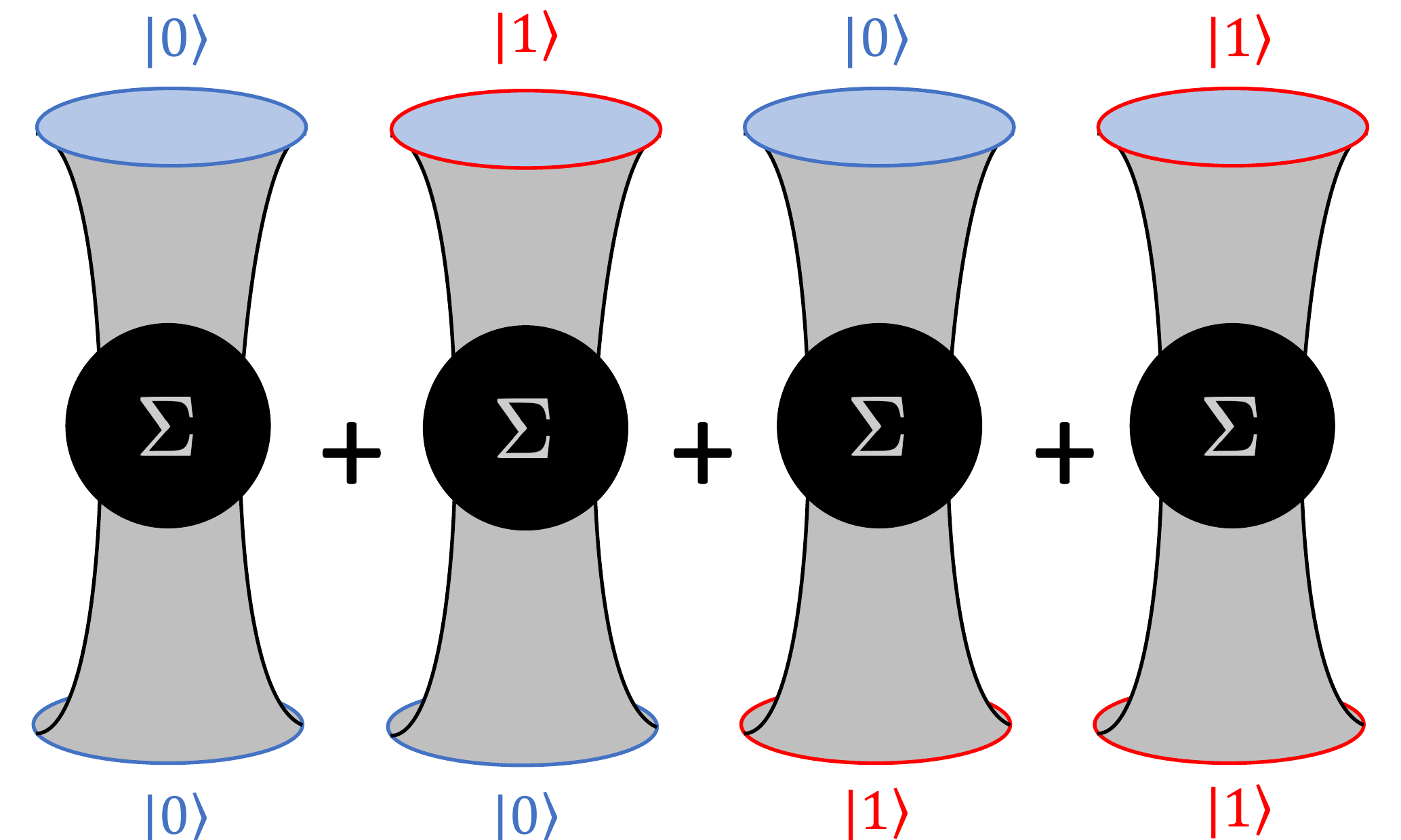}
\end{matrix}\nonumber \\
&=& \begin{matrix}
\includegraphics[height=2.7cm]{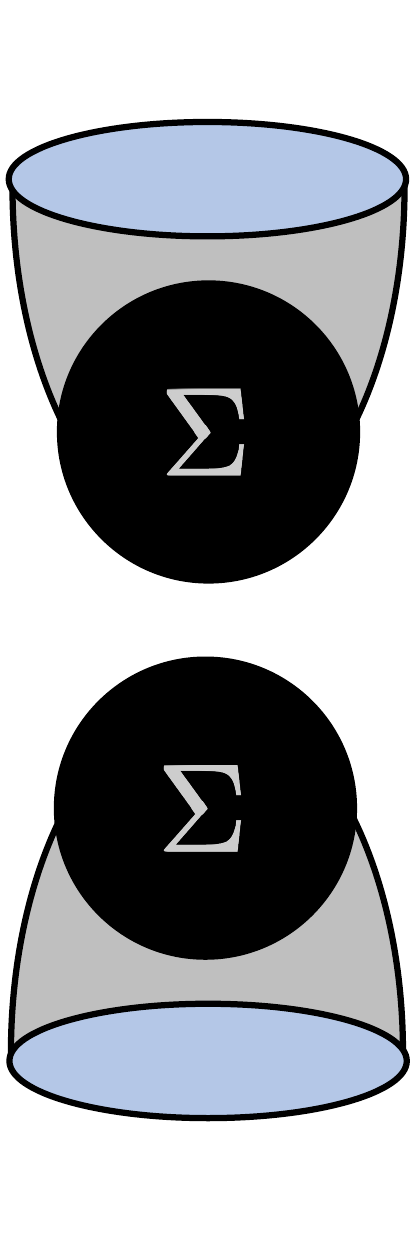}
\end{matrix}+ \begin{matrix}
\includegraphics[height=2.7cm]{pic/factwh2}
\end{matrix}%\nonumber\\
\label{sum0}
\eea
There is no way to use the overlap between different vacua to represent the disconnected geometries, so we need to keep those contributions and use black circles to denote the superposition of different vacua. As for the four connected geometries, if they cancel each other, we would end with a factorized partition function
\bea\label{Zfa}
Z\left[ \bigcirc\cup  \bigcirc \right]
=\begin{matrix}
\includegraphics[height=2.8cm]{pic/fa1}
\end{matrix}\,.
\eea
Then, we can say we get 
$Z\left[ \bigcirc\cup  \bigcirc \right]=Z\left[ \bigcirc \right]\times Z\left[ \bigcirc \right]$.

%\newpage
\subsection{Cancellation between wormholes}

The main task for the remainder of this section is to bootstrap the relation between the connected geometries as shown in \eqref{sum0} by requiring no factorization puzzle. %By no factorization puzzle, we actually mean the bulk connecting geometries canceling each other. 
We will get the expression of $S$ such that all the bulk geometry that connects different boundaries sum over to give a zero.

Firstly, let's try some simple cases. For the ($n=2$, $k=2$) case, the partition function can be illustrated as
\bea \label{Zfa0}
Z\left[ \bigcirc\cup  \bigcirc \right]
&=& \begin{matrix}
\includegraphics[height=2.7cm]{pic/fa1}
\end{matrix}
+\begin{matrix}
\includegraphics[height=2.7cm]{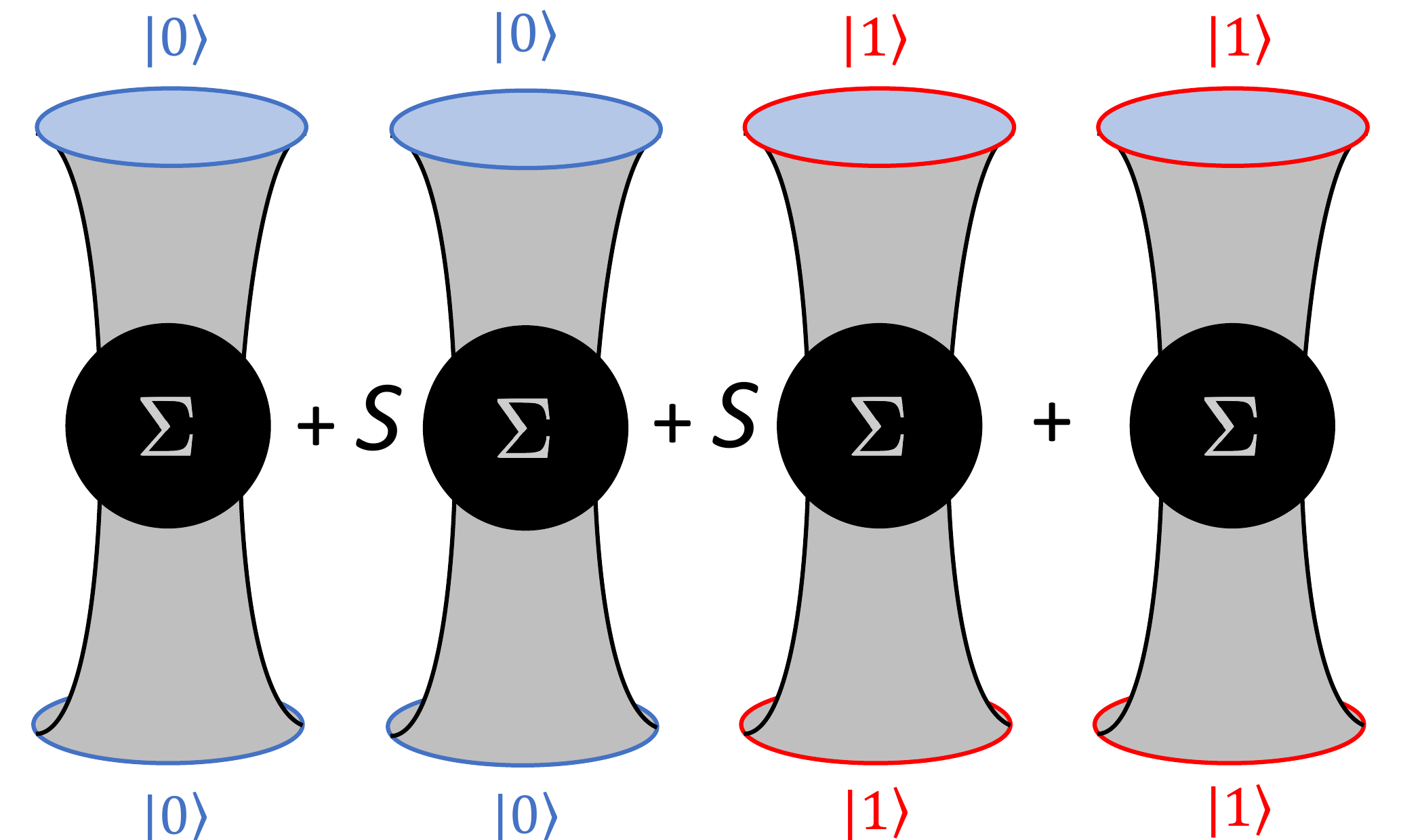}
\end{matrix}\,.
\eea
The answer seems very obvious. If $S=-1$, the geometries connecting different boundaries cancel each other, and we get a factorized result for $Z\left[ \bigcirc\cup  \bigcirc \right]$. 
Moreover, $S=-1$ preserves the group structure of $\ZZ_2$, because as argued before we want $S^2=1$ such that the group is cyclic.

For the $(n=3$, $k=2)$ case, we have three boundaries and two degenerate vacua in the theory. 
We are going to fill all possible geometries between those three boundaries and calculate $Z\left[ \bigcirc\cup  \bigcirc \cup  \bigcirc\right]$. The geometry can be represented as follows,
\bea
&Z& \left[ \bigcirc\cup  \bigcirc \cup  \bigcirc\right]\nonumber\\
&=& \begin{matrix}
\includegraphics[height=1.8cm]{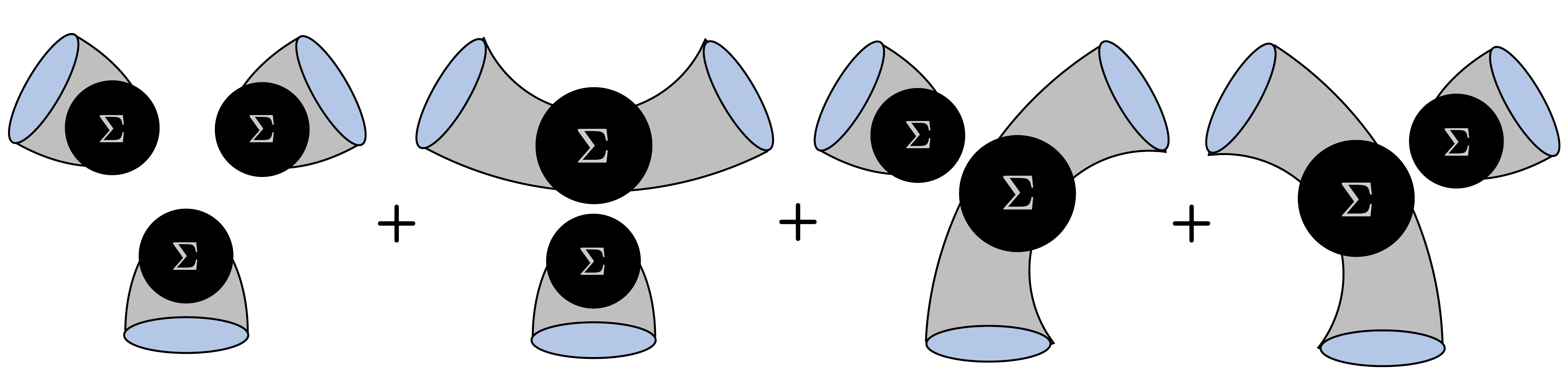}
\end{matrix}\nonumber\\
&+& \begin{matrix}
\includegraphics[height=1.8cm]{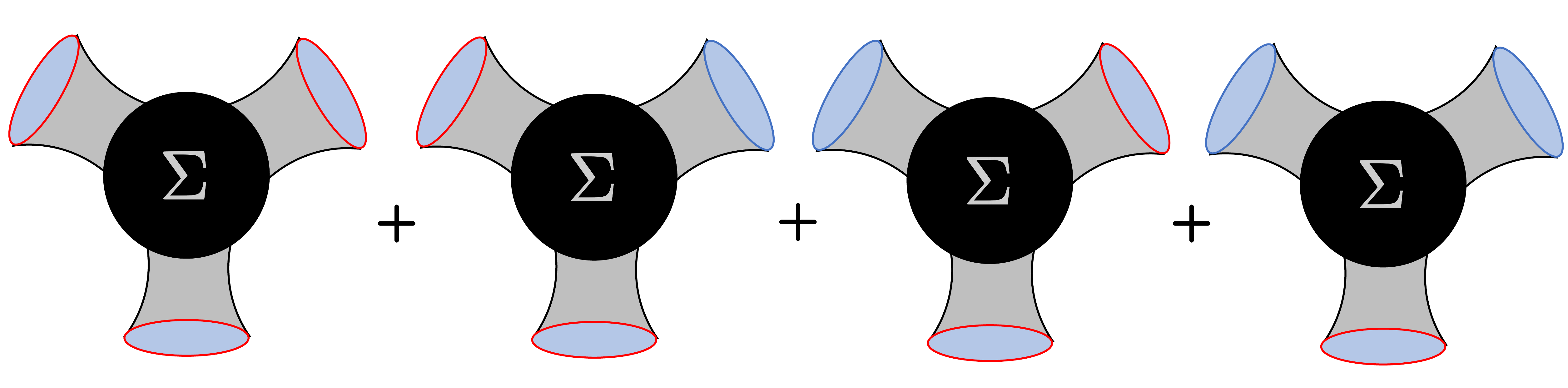}
\end{matrix}\nonumber\\
&+& \begin{matrix}
\includegraphics[height=1.8cm]{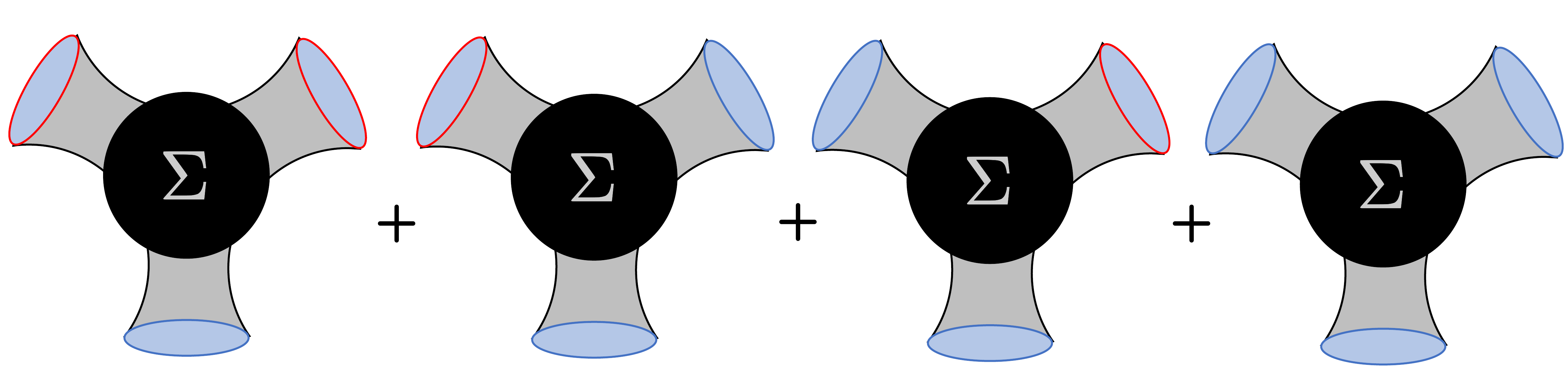}
\end{matrix}\,.\label{23}
\eea
For cases with $n\neq 2$, since we only want completely disconnected geometries, all the partially connected geometries should also be canceled. Say it differently, all the connected geometries with $m<n$ boundaries should cancel each other before the cancellation works for the $n$-boundary case. 
The last three diagrams in the first line of \eqref{23} all contain wormholes connecting two boundaries. For those diagrams to be individually canceled, we need to impose the same constraint as the ($n=2$, $k=2$) case. So with constraints 
\be\label{2cons}
S+1=0\,, ~~~\text{and}~~~ S^2=1\,,
\ee
the wormholes connecting two boundaries vanish.
To cancel the geometries connecting three boundaries, we need to ask
\be\label{3cons}
S^2+2S+1=(S+1)^2=0\,, ~~~\text{and}~~~ S^2=1\,.
\ee
There is a solution of constraints \eqref{2cons} and \eqref{3cons}, which is $S=-1$. The $(n=3$, $k=2)$ case is same as the $(n=2$, $k=2)$ case. 

Note that we have required bulk geometries connecting the same number of boundaries to vanish step by step. In the $(n=3$, $k=2)$ case, we first asked the wormholes connecting two boundaries to vanish and then asked the geometries connecting three boundaries to give out a zero answer. It seems like nothing prevents partially and fully connected geometries conspire altogether to produce a vanishing answer. However, a partially connected geometry can never cancel a fully connected one, because they have different gravitational partition functions. For example, in the $(n=3$, $k=2)$ case, a partially connected geometry has a partition function proportional to $Z_1\times Z_2$, while a fully connected geometry would be proportional to $Z_3$. So the constraints listed in \eqref{2cons} and \eqref{3cons} should respect the hierarchy of connected boundaries.

For the $(n=2$, $k=3)$ case, the three degenerate vacua are denoted as $\{\ket{0},\ket{1},\ket{2}\}$, represented by blue, red, and green colors respectively. %We also use three different colors blue, red, and green respectively to represent those vacua.
The bulk partition function can be written as
\bea\label{n2k3}
&Z& \left[ \bigcirc\cup  \bigcirc\right]\nonumber\\
%&=& \begin{matrix}
%\includegraphics[height=3.3cm]{3k1}
%\end{matrix}\nonumber\\
%&+& \begin{matrix}
%\includegraphics[height=3.3cm]{3k2}
%\end{matrix}\nonumber\\
&=& \begin{matrix}
\includegraphics[height=2.5cm]{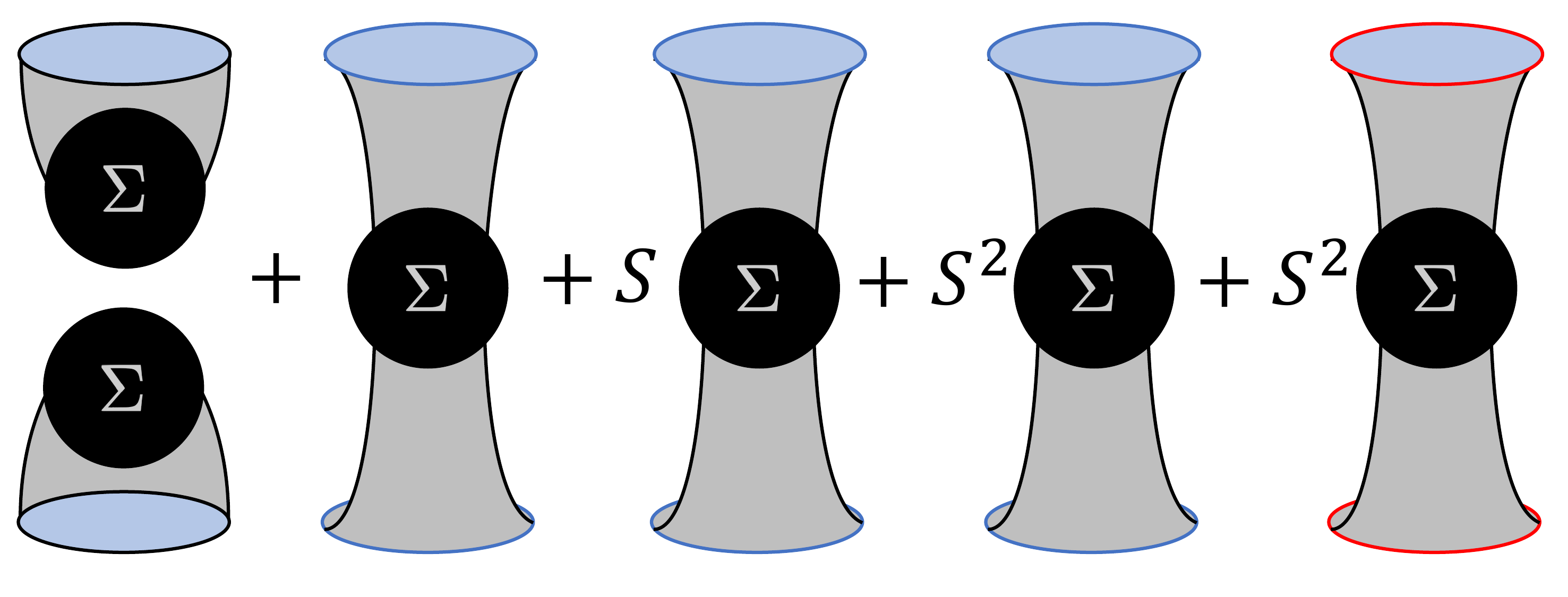}
\end{matrix}\nonumber\\
&+&\begin{matrix}
\includegraphics[height=2.5cm]{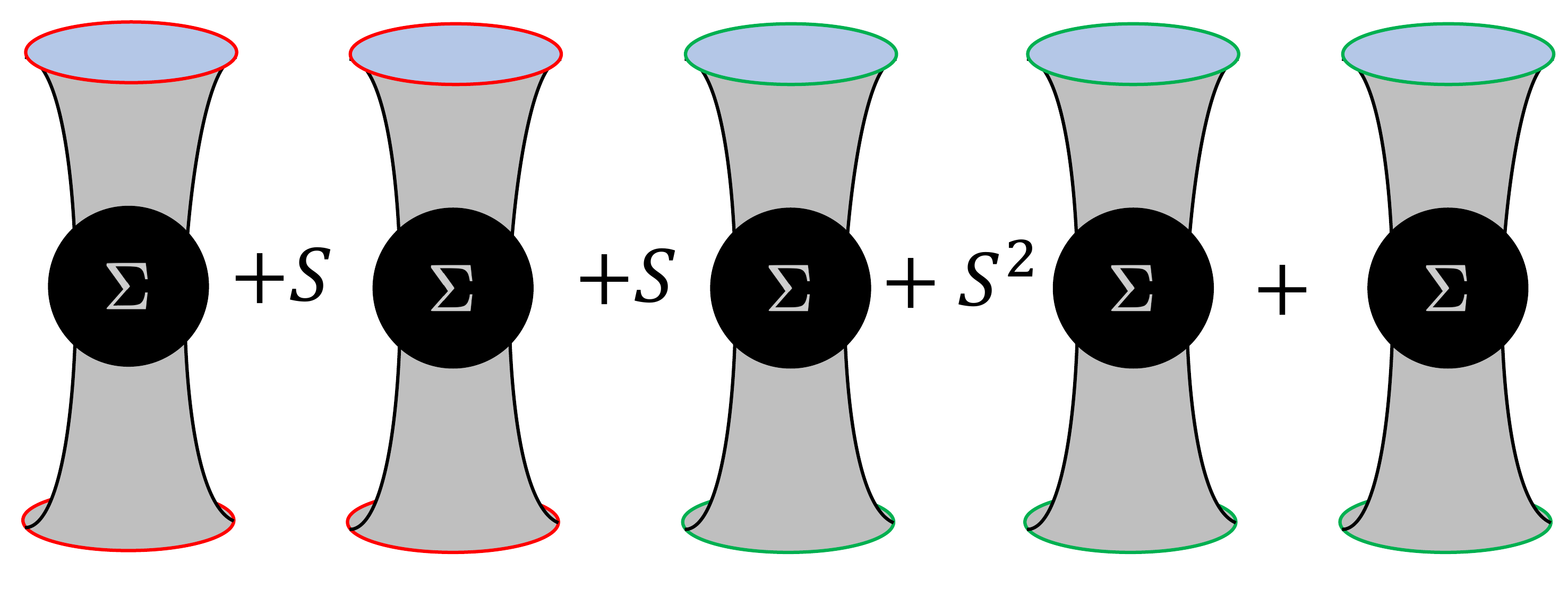}
\end{matrix}
\eea
The bulk cancellation works if the following constraints are satisfied
\bea\label{3kcons}
& S^2+S+1=0\,,\nonumber\\
& S^3=1\,.
\eea
The solution can be easily solved, which gives us
\be\label{S3}
S=\exp\left[\frac{2\pi i}{3}\right]\,.
\ee
So $S=-1$ doesn't work generally, we want a general result for $n$ boundaries and $k$ degenerate vacua.

It is natural to ask why don't we demand partition function canceling between the different colors as opposed to individually. For example, is it possible for the third diagram in the first line of \eqref{n2k3} to cancel the second diagram in the second line?
Actually, those two diagrams are the same. 
Say it differently, rotating the whole figure $(a)$ in \eqref{3WH} is a diffeomorphism (gauge transformation). So asking two times a diagram to be zero is the same as demanding the diagram equals zero, and there is no non-trivial solution.
So, the constraints shown in \eqref{3kcons} gotten by demanding the diagrams cancel each other is the only natural choice. The configurations with the same colors can be regarded as like terms and the cancellation works only after uniting like terms.
Certainly, for more general $n$ and $k$, there can be other ways to twist the diagrams and get other combinations of figures. But if twisting the up trumpet brings in an extra $S$, twisting the down trumpet in the same direction brings in a $S^*$. One can manipulate the diagrams however one wants.
After all, all the other combinations give the same equations as shown in \eqref{3kcons}, if not the trivial ones.

~\\ \noindent \textbf{General case}~\\

What we have learned from the simple cases is that to calculate the $n$ case, the wormholes connecting $m < n$ boundaries should also cancel each other. 
So the factor depends on $k$ but should not depend on $n$. 
Let us look at the $n=2$ case with $k$ degenerate vacua.
The $k$ vacua can be labeled by $\{\ket{0},\ket{1},\cdots,\ket{k-1}\}$. We are going to name the two boundaries as up and down boundaries. There are $k$ different choices for the down boundary, and we would like to see how many choices for the up vacuum with the fixed down vacuum. Comparing the up and down vacuum gives out the extra factors whose powers are determined by the difference between the vacua. Supposing the down boundary is in $\ket{\alpha}$ vacuum, the constraint from the cancellation can be written as
\be
\sum_{\alpha=0}^{k-1}\sum_{i=0}^{k-1}\begin{matrix}
\includegraphics[height=2.8cm]{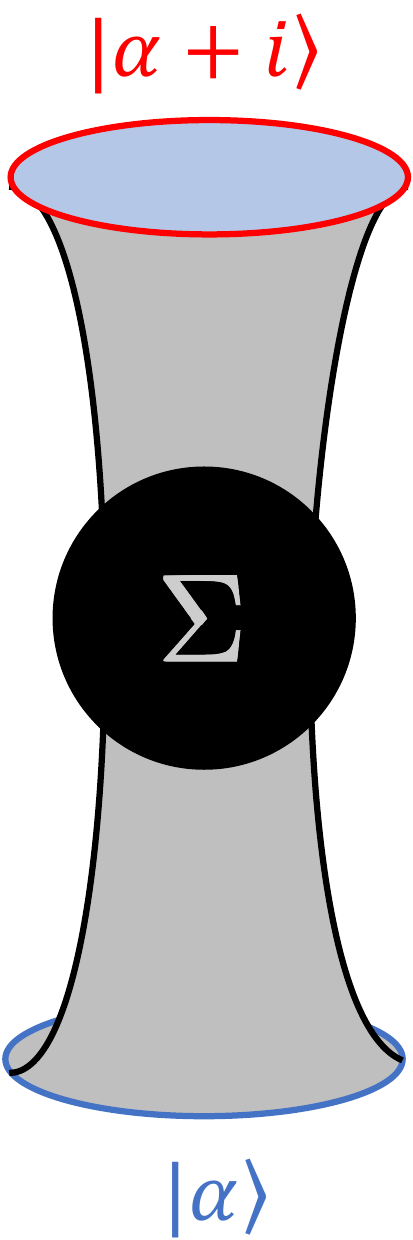}
\end{matrix}~~
= ~~\sum_{\alpha=0}^{k-1}\sum_{i=0}^{k-1}~S^{i}\times ~~\begin{matrix}
\includegraphics[height=2.8cm]{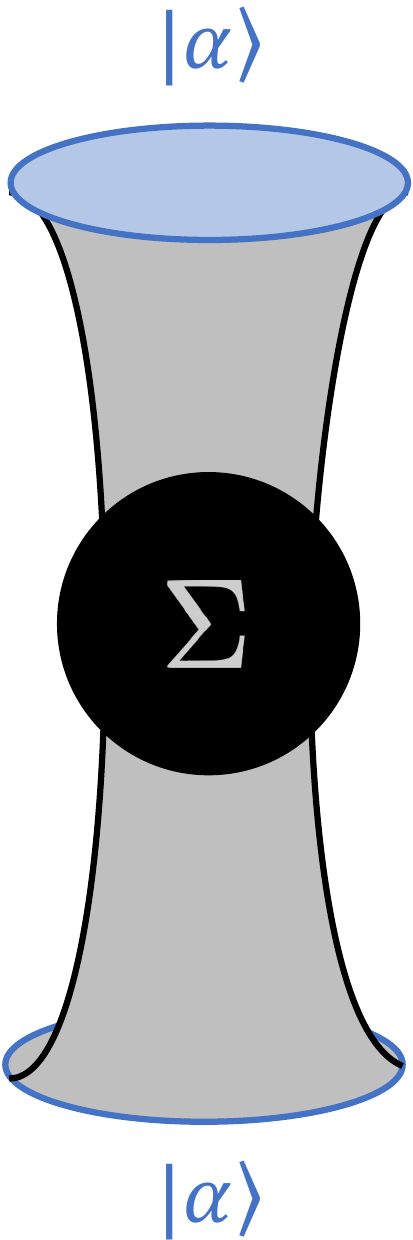}
\end{matrix}=0\,,
\ee
We want the cancellation works for every choice of down vacuum $\ket{\alpha}$ such that all the connected diagrams sum over to zero. Or equivalently, we can rewrite the above expression as
\be
\sum_{i=0}^{k-1}~S^{i}\times~~ \sum_{\alpha=0}^{k-1}~~\begin{matrix}
\includegraphics[height=2.8cm]{pic/gen2}
\end{matrix}=0\,,
\ee
So the constraint from the bulk connected geometry cancellation is
\be\label{cons1}
\sum_{i=0}^{k-1}~S^{i}=0\,.
\ee
We also have another constraint from the cyclic group structure
\be\label{cons2}
S^{k}=1\,.
\ee
The solution of the constraints \eqref{cons1} and \eqref{cons2} can be easily solved, which is
\be
S=\exp\left[\frac{2\pi i}{k}\right]\,.
\ee

\begin{figure*}
  \begin{center}
  	\includegraphics[width=12cm]{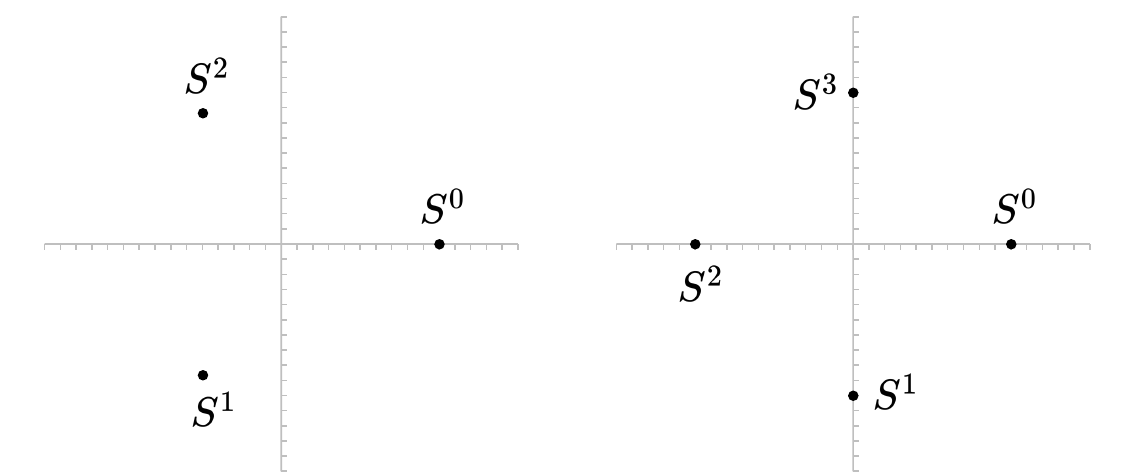}
  \end{center}
  \caption{The distribution of extra factor $S$ on the complex plane. The left figure is the case $k=3$, while the $k=4$ case is shown in the right figure.}\label{SSS}
\end{figure*}

For the geometry connecting $n$ boundaries, we can fix one of the boundaries as $\ket{\alpha}$, and then compare $\ket{\alpha}$ with the other $n-1$ legs. So we are going to compare $\ket{\alpha}$ with those legs one by one, which gives out a product of $n-1$ factors. The diagram can be expressed as
\bea
&~& \overbrace{\sum_{i=0}^{k-1}~S^{i}\times\cdots\times\sum_{j=0}^{k-1}~S^{j}}^{\text{product of}~n-1~\text{factors}}\times~~ \sum_{\alpha=0}^{k-1}~~\begin{matrix}
\includegraphics[height=2.5cm]{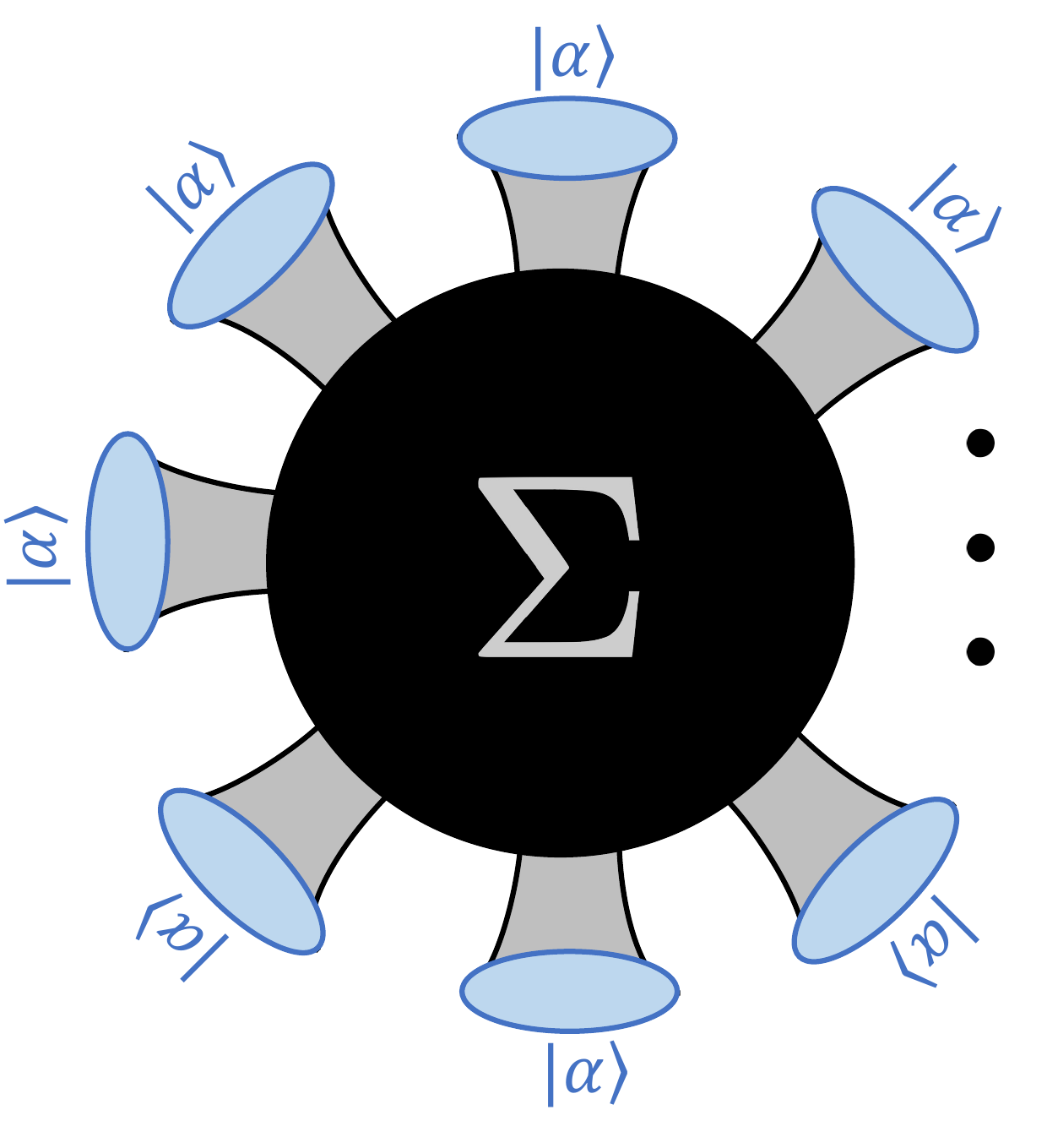}
\end{matrix}\nonumber\\
&~& =
\left(\sum_{i=0}^{k-1}~S^{i}\right)^{n-1}\times~~ \sum_{\alpha=0}^{k-1}~~\begin{matrix}
\includegraphics[height=2.5cm]{pic/gen3}
\end{matrix}=0\,.\nonumber\\
\eea
The constraint from the above diagram is
\bea
\left(\sum_{i=0}^{k-1}S^i\right)^{n-1}=0~~ \iff~~ \sum_{i=0}^{k-1}S^i=0\,.
\eea
The expression is indeed independent of $n$, so this equation is satisfied, all the cases with $m < n$ boundaries are also satisfied.
Just like the $n=2$ case, the expression for factor $S$ is
\be\label{expressionS}
S=\exp\left[\frac{2\pi i}{k}\right]\,,
\ee
which also respects the group structure.
It is easy to show that simple cases \eqref{3cons} and \eqref{S3} can also be checked, which are
\bea
\exp\left[\frac{2\pi i}{k}\right]_{k=2}=-1\,,~~~\exp\left[\frac{2\pi i}{k}\right]_{k=3}=\exp\left[\frac{2\pi i}{3}\right]\,.
\eea
The factor $S$ can be illustrated on the complex plane shown in Fig. \ref{SSS}. We have the $k=3$ case shown on the left hand of the figure and the $k=4$ case is shown next to it. 

Note that it doesn't matter to assign $S$ or $S^{*}$ to the $i+1$ vacuum. The cancellation works because we need to compare one specific vacuum $\ket{i}$ with every other vacuum and all the vacua are uniformly distributed on a circle centered at the origin on the complex plane.
Take the $n=2,k=3$ case as an example, here we defined
\begin{equation}
	\ket{1}=S\times \ket{0}\,,~~~~\ket{2}=S\times \ket{1}\,,~~~~\ket{0}=S\times \ket{2}\,.
\end{equation}
The final result would be the same if we had used a different definition 
\begin{equation}\label{permu}
	\ket{2}=S\times \ket{0}\,,~~~~\ket{1}=S\times \ket{2}\,,~~~~\ket{0}=S\times \ket{1}\,.
\end{equation}
On the complex plane, the above definitions correspond to a different order of points on the first figure shown in Fig. \ref{SSS}. The story is invariant under the permutation of different vacua.

As a brief summary of the section, we have shown that by adding degenerate vacua, the bulk partition function is indeed factorized because of the cancellation of the bulk connecting geometries. 
We assumed $k$ degenerate vacua in the theory, and bulk geometries connecting the boundaries in different vacua give out extra factors $F(S)$. The expression of factors is solved from the constraints \eqref{cons1} and \eqref{cons2}. 
It is easy to see that the bulk geometries connecting different boundaries cancel each other, and we end up with a factorized partition function.

It is important to note that unlike some schemes trying to handle the factorization puzzle by completely killing the bulk topological modes, the bulk topological modes are protected in our prescription to realize factorization. This can be seen from \eqref{Zfa}, where $\Sigma$ represents the summation of handlebodies.
Moreover, the calculation doesn't mean the wormhole contributions are completely killed, we just want to show that the wormhole contributions add up to zero. 
When a specific boundary vacuum is chosen, we can still see the wormholes.

Note that the twist is characterized by a continuous U(1) parameter, while we have used a discrete $\ZZ_k$ symmetry to demonstrate the mechanism of realizing factorization. The reason for such a choice is that $\ZZ_k$ symmetry is good enough to capture the essence of the twist shown in \eqref{3WH}, and the factorization can be related using this relatively simple version of U(1) symmetry. 
The $k\to \infty$ limit of $\ZZ_k$ can be regarded as the U(1) symmetry. The mechanism demonstrated in this paper should work for the $k\to \infty$ limit, so we have enough reason to believe the factorization puzzle can be understood if the symmetry is the continuous U(1) symmetry. Nevertheless, an explicit demonstration of the wormhole cancellation with continuous U(1) symmetry is worth careful research, and we will leave it for further studies.

%%%%%%%%%%%%%%%%%%%%%%%%%%%%%%%%%%%%%%%%%%%%%%%%%%%%%%%%%%%%%%%%%%%%%%%%%%%%%%%%%%%%%%%%%%%%%%%%%%%%

%\newpage
%\input{sec/4-SYK.tex}

\section{Cancellation between wormhole-type correlations in the SYK model}
\label{dual}

In this section, we are going to study the cancellation between wormhole-type correlations in the SYK model, from which we can see the wormhole cancellation is because of the associated extra complex phase.

The cancellation being demonstrated in this section is largely inspired by the previous section. We will discuss further relations between the two ways of cancellation in the discussion section. The JT gravity discussed before is a two-dimensional topological theory, whose boundary theory is described by the Schwarzian action \eqref{Schwarzian}.
It was shown that the low energy dynamics of the SYK model \cite{Sachdev1993,Kitaev2015,Kitaev2017} can also be described by the Schwarzian theory \cite{Maldacena2016a,Kitaev2017}. 
The JT gravity can be regarded as the low energy limit of the SYK model. So the SYK model provides a good model for understanding problems in JT gravity. 
In this section, we would adopt an even simpler toy model, i.e. one-time-point SYK model \cite{Saad2021}, where the time contour is replaced by a single instant of time.

%In general, the couplings in the SYK model follow a specific distribution, the Gaussian distribution as an example. 
The factorization puzzle is not necessarily a puzzle in the averaged theory, because the wormholes can be regarded as the variance. 
So, to consider the factorization puzzle, we are mainly interested in the SYK model with fixed couplings because a specific boundary theory rather than an ensemble of theories matches our general understanding of holography.
It was shown by the authors of \cite{Saad2021} that when the couplings are fixed, the wormhole saddles persist.
Moreover, for the case with more than two fermions, the wormhole saddles are not sensitive to whether we average over the couplings or not, i.e. they are self-averaging saddles.
The main task of the section is to see if a mechanism for wormhole cancellation appears in the one-time-point SYK model with fixed couplings. We will see the mechanism is similar to the one discussed in the previous section.

\subsection{One-time-point SYK model with fixed couplings}
\label{otpSYK}

First of all, let us review the basics of the one-time-point SYK model, following paper \cite{Saad2021}.
The one-time-point SYK model is a theory of $N$ Grassmann numbers, with partition function
\be
Z=\int \dd^N \psi~ \exp{\left[ i^{q/2}\sum_{1<i_1<\cdots<i_q<N} J_{i_1\cdots i_q}\psi_{i_1}\psi_{i_2}\cdots\psi_{i_q}\right]}\,.
\ee
We can denote the product of Grassmann numbers as $\psi_{i_1\cdots i_q}=\psi_{i_1}\cdots\psi_{i_q}$. The couplings $J_{i_1\cdots i_q}$ are Gaussian distributed, such that the mean and variance can be written as
\begin{align}
	\bracket{J_{i_1\cdots i_q}} &= 0\,,\nonumber
	\\
	\bracket{J_{i_1\cdots i_q}J_{j_1\cdots j_q}} &= \frac{J^2(q-1)!}{N^{q-1}}\delta_{i_1 j_1}\cdots\delta_{i_q j_q}\,,
\end{align}
for constant $J$ which can be taken to 1. Here $q$ is taken to be an even number greater than two. %Now, we can denote the variance as $\#={(q-1)!}/{N^{q-1}}$.

\begin{figure}
  \begin{center}
  	\includegraphics[height=3cm]{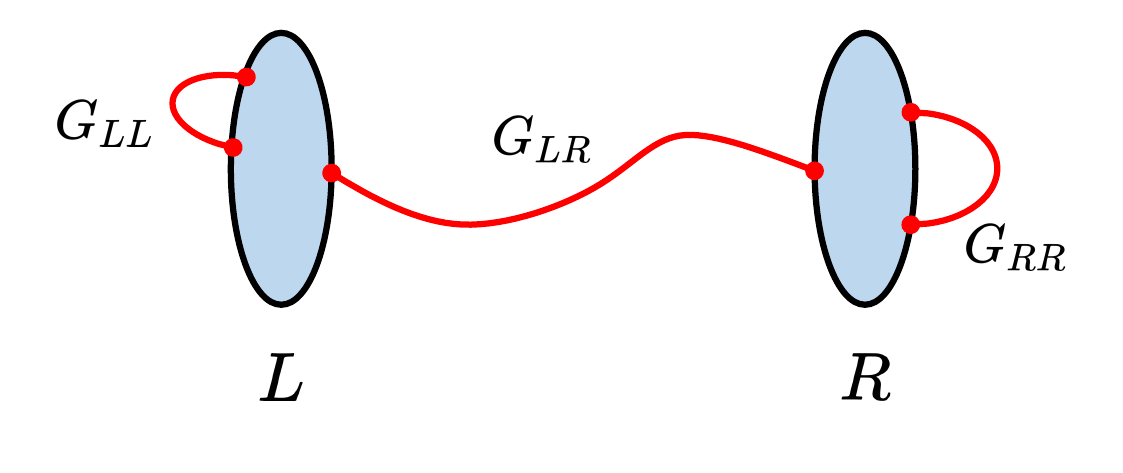}
  \end{center}
  \caption{Sketch of collective fields $G_{LL}$, $G_{RR}$, and $G_{LR}$ in the SYK model. While we only have wormhole-type correlation $G_{LR}$ in the one-time-point SYK model.}\label{GGG}
\end{figure}

What are wormholes in the SYK model? We can replicate the system and calculate the correlation between the left and the right systems. The whole model can be rewritten with the collective fields $\Sigma$ and $G$. As shown in Fig. \ref{GGG}, the correlation between the two replicas $G_{LR}$ can be interpreted as the wormhole. 
One advantage of the one-time-point SYK model, as we will see, is that we only have wormhole-like correlations $G_{LR}$ and $\Sigma_{LR}$ in the theory. $G_{LR}$ is defined as 
\be\label{GLR}
G_{LR}=\frac{1}{N}\sum_{i=1}^N\psi^L_i\psi^R_i\equiv \frac{1}{N}\psi^L_i\psi^R_i\,,
\ee
and $\Sigma_{LR}$ is the Lagrangian multiplier that enforces \eqref{GLR} in path integral. $G_{LL}$ and $G_{RR}$ are zero because there is only one point on each boundary and the square of a Grassmann number vanishes. We will briefly discuss a more general model in the next subsection and see how the conclusions in this section can be extrapolated to the regular SYK model.
The collective fields can be easily defined when we average the theories over coupling constants $J_{j_1\cdots j_q}$.
However, to consider the factorization puzzle, one should begin with a version of the SYK model with fixed couplings, because factorization is not necessarily a property in the averaged theory. The puzzle itself can be reformulated as the existence of bulk-connected wormholes in the non-averaged theory.

\begin{figure}
  \begin{center}
  	\includegraphics[height=6cm]{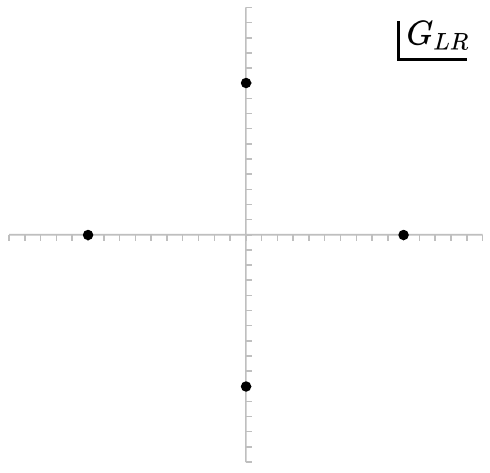}
  \end{center}
  \caption{Wormhole saddle points in $Z_L Z_R$ for the $q=4$ case.}\label{q4}
\end{figure}

We are mainly interested in if the non-averaged partition function $Z_L Z_R$ 
\bea
&~& Z_L Z_R = \int\dd^{2N}\psi \nonumber\\&\times&
\exp \Big{[} i^{q/2} \sum_{1<i_1<\cdots<i_q<N} J_{i_1\cdots i_q}\left(\psi^L_{i_1 \cdots i_q}+\psi^R_{i_1 \cdots i_q}\right)\Big{]}\,\nonumber\\
\eea
can be factorized or not.
It can be shown that, in the large $N$ limit, 
\bea
&~& Z_LZ_R\Big{|}_{\text{wormholes}} = \bracket{Z_LZ_R}_J\Big{|}_{\text{wormholes}}\nonumber\\
&=& \frac{N}{2\pi}\int\dd g_{LR}\int\dd\sigma_{LR}\nonumber\\
&~& \times\exp[ N(\log(ie^{-\frac{i\pi}{q}}\sigma_{LR})-i\sigma_{LR}g_{LR}-\frac{1}{q}g_{LR}^q) ]\Big{|}_{\text{wormholes}}\,,\nonumber\\
\eea
and the main contributions are from the wormhole saddle points.
See Appendix \ref{BB} for more details.
$g_{LR}$ and $\sigma_{LR}$ are just collective fields
\be
\Sigma_{LR}=ie^{-i\frac{\pi}{q}}\sigma_{LR}\,,~~~~~~G_{LR}=e^{i\frac{\pi}{q}}g_{LR}\,.
\ee
and we can get the $q$ solutions of the saddle point equation \eqref{saddleE}, which are
\bea\label{solgs0}
g^{m}_{LR} &=& e^{-i\pi/q}e^{-\frac{2\pi m i}{q}}\,,\\
\sigma^{m}_{LR} &=& -i e^{i\pi/q}e^{\frac{2\pi m i}{q}}\,,
\eea
and label them by $0\leq m\leq q-1$. In Fig. \ref{q4}, the solutions are labeled on the complex plane for the $q=4$ case.
Those wormhole saddle points are largely reminiscent of the bulk wormhole distributions shown in Fig. \ref{SSS}. So we can say that we found another support of our bulk discussion in the SYK model.
 
To further check the cancellation, we can calculate the partition function $Z_LZ_R$ in large $N$ limit. 
With the wormhole solutions \eqref{solgs0} at hand, the partition function $Z_LZ_R$ can be worked out as the summation of all the saddle points
\bea
Z_L &Z_R&\Big{|}_{\text{wormholes}}
%\nonumber\\
%&\approx & \frac{N}{2\pi} \sum_{m=0}^{q-1} \Big{[}(ie^{-\frac{i\pi}{q}}\sigma_{LR})^N \nonumber\\
%&\times&\exp[N(-\sigma_{LR}g_{LR}-\frac{1}{q}g_{LR}^q) ]\Big{|}_{g_{LR}=g^{m}_{LR},\sigma_{LR}=\sigma^{m}_{LR} }\Big{]}\,\nonumber\\
%&\propto &
\propto \sum_{m=0}^{q-1} e^{\frac{2\pi m i}{q}N}e^{-N(1-\frac{1}{q})}\,.\label{15}
\eea
Note that the $q$ wormhole saddle points sum together to zero when $N$ is not a multiple of $q$. 
In general, there is no obvious relation between $N$ and $q$. 
So we conclude that the wormholes associated with extra phase can cancel each other in the partition function. Moreover, the distribution of the wormhole solution matches the bulk calculation demonstrated in the previous section.
It is worth noting that when $N$ is divisible by $q$, the above cancellation doesn't work and further studies are needed to understand this specific situation.
Or one may say that in order for the cancellation, $N$ should not be divisible by $q$, otherwise we need extra novel mechanisms to cancel the wormholes.

\subsection{Turning on $G_{LL}$ and $G_{RR}$}

The above demonstration can be extrapolated to a more general setting of the SYK model with fixed couplings. We can turn on time dependence on the $L$ and $R$ replicas. The consequences of the boundary time dependence are nontrivial $G_{LL}(t,t')$ and $G_{RR}(t,t')$ defined as
\bea
G_{LL}(t,t')=\frac{1}{N}\psi_i^{L}(t)\psi_i^{L}(t')\,,\\
G_{RR}(t,t')=\frac{1}{N}\psi_i^{R}(t)\psi_i^{R}(t')\,,
\eea
as well as nontrivial $\Sigma_{LL}$ and $\Sigma_{RR}$, and we would also have more $G_{LR}$ and $\Sigma_{LR}$ configurations with different $(t,t')$, as shown in Fig. \ref{GGG}.

Although a trivial partition function for $G_{LR}$ in the one-time-point SYK model with fixed couplings already means factorization between two boundaries, turning on the time dependence on the left and right boundaries can help us see the factorization between the two boundaries more clearly. 
As discussed at the beginning of the section, the factorization puzzle is only necessary to be addressed in the SYK model with fixed couplings. We do not want to average over couplings and would like to work with fixed couplings to address the factorization puzzle, because a specific boundary theory rather than an ensemble of theories matches our general understanding of holography.

%or correspondingly, we can add $G_{LL}$ and $G_{RR}$ back to the theory

When we only have a single instant of time on each boundary, we only have $G_{LR}$ which represents the correlations between different boundaries, i.e. wormholes in the theory. 
In the more general setting, $G_{LL}$ and $G_{RR}$ are not zero, and these bilocal fields, together with $\Sigma_{LL}$ and $\Sigma_{RR}$, can be used as the equivalent description of boundary theories on each boundary. 
Let us only consider the saddle point contributions in partition function $Z^2$ with fixed couplings, the partition function can be written as the summation of contributions of different saddles. Note that we can have disconnected saddles containing $G_{LL}(t,t')$ and $G_{RR}(t,t')$ on each boundary and connected wormhole saddles. The disconnected saddle contributions can be factorized while the wormholes can not. 
With fixed couplings, $G_{LR}$ has a similar behavior as in the one-time-point SYK model. Each $G_{LR}$ gives out a non-zero contribution to $Z^2$, and $q$ of them together give zero contribution to the partition function. 
The partition function $Z^2$ can be written as 
\bea
Z^2&=&\sum_{\text{saddles}}e^{-I}\nonumber\\
&=&\sum_{\text{disconnected saddles}}e^{-I}+\sum_{\text{wormholes}}e^{-I|_{\text{wormholes}}}\,,\nonumber\\
\eea
The wormhole contributions sum over to zero and we are only left with disconnected saddle contributions. There is no factorization puzzle for disconnected saddles contributions.
$Z_L Z_R$ equals zero in the one-time-point SYK model means the wormhole-like correlation cancels each other in the general SYK model with fixed couplings and the partition function can be factorized into left and right parts, where each part only contains $G_{LL}$ and $G_{RR}$.

So as a brief summary, for general $N$ and $q$, the wormhole saddles defer each other with phases
\be\label{SYKfactor}
S=\exp\left[\frac{2\pi i}{q}m\right]\,,~~~\text{with}~~~0\leq m\leq q-1\,.
\ee
We saw a similar behavior in the JT gravity.
The wormhole saddles for the $q=4$ case are shown in Fig. \ref{q4}, where we have four wormhole saddles in the path integral each with different phases. The four saddles \eqref{solgs0} form a cyclic group $\ZZ_4$ which is reminiscent of the vacua structure discussed in the previous section and the distribution shown in Fig. \ref{SSS}.%as can be seen from Fig. \ref{q4}. 
Note that from the partition function shown in \eqref{15}, the cancellation works when $N$ is an integral multiple of $q$.
So another conclusion of this section is that factorization puts constraints on the theory. In the current case, $N$ should not be divisible by $q$, otherwise we need extra novel mechanisms to cancel the wormholes.
%%%%%%%%%%%%%%%%%%%%%%%%%%%%%%%%%%%%%%%%%%%%%%%%%%%%%%%%%%%%%%%%%%%%%%%%%%%%%%%%%%%%%%%%%%%%%%%%%%%%

%\newpage
%\input{sec/5-summary.tex}

\section{Conclusion and discussion}
\label{conc}

In this paper, we demonstrate two examples where 
the cancellation between bulk connecting geometries can be realized thus the factorization puzzle can be avoided. 
%The structure might not depend on a specific theory, but if factorization is regarded as the essential property, the structure should be a constraint on the general theory. 
If factorization of bulk gravity dual to decoupled boundary theories is regarded as the essential property, 
the examples might provide some important clues for the general structure a bulk theory should associated with.
In general, we propose that the wormholes connecting different boundaries should be divided into different sectors, and those sectors are associated with different phases on the complex plane. 
With the extra complex phase, the bulk connecting geometries can cancel each other in the partition function, which results in a factorized result. 
%The cancellations are shown both in the JT gravity and the SYK model.

In the JT gravity, by allowing degenerate vacua with a specific structure, the bulk geometries connecting different asymptotic boundaries are associated with extra factors. The expression of the factors can be bootstrapped by requiring the cancellation between those bulk connecting geometries. As a result, the extra factors are complex phases and there should be several different sectors of those geometries. The distributions of the sectors are shown on the complex plane in Fig. \ref{SSS}. The connecting geometries cancel each other we end up with a factorized gravity partition function, which matches the boundary partition function.

It is well-known that the boundary theory of the JT gravity, i.e. the Schwarzain theory, is the low energy limit of the SYK model. 
We adopt a very simple toy model, the one-time-point SYK model, to study the behavior of the wormhole-type correlation $G_{LR}$. 
In the path integral calculation of two replica quantities, we find that there are $q$ different wormhole saddles in the saddle point approximation.
The distribution of the $q$ wormhole saddles is super similar to the distribution shown in Fig. \ref{SSS}. 
Those saddles are also associated with different phases and cancel each other in the path integral. 
Although the distribution of the saddles doesn't depend on the relation between $N$ and $q$, $N$ is not an integral multiple of $q$ is vital for the cancellation.
So if one claims the factorization is an essential property of a theory and no extra novel ingredients should be added to cancel the wormholes, the factorization may put a constraint on the theory, i.e. $N$ should not be an integral multiple of $q$.

~\\ \noindent \textbf{Discussion:}~\\

\begin{figure*}
\bea
 Z\left[ {\color{blue} \bigcirc} \cup  {\color{red} \bigcirc }\cup  {\color{green} \bigcirc}\right]=%\nonumber\\~\nonumber\\&~&
 \begin{matrix}
 	 	\includegraphics[height=2.8cm]{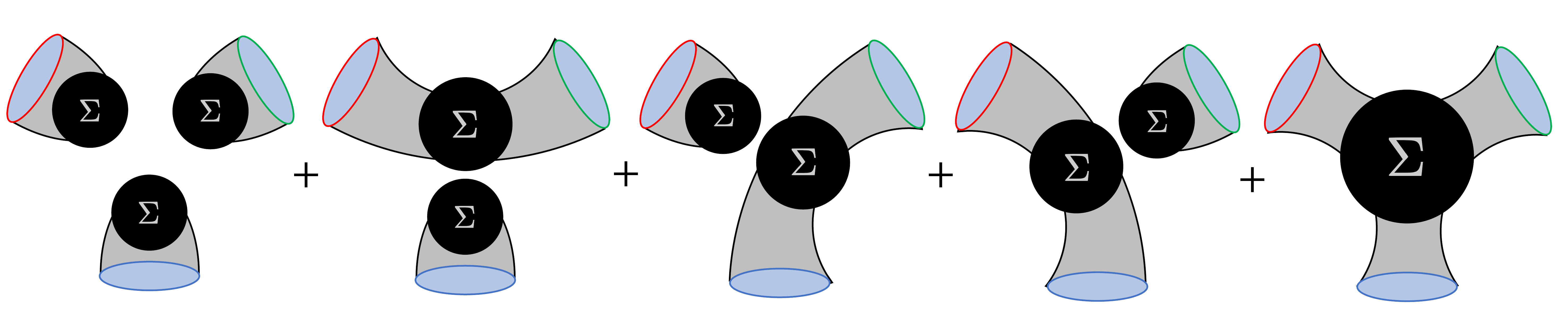}
 \end{matrix} \nonumber
\eea
  \caption{Bulk geometries specific boundary vacua are chosen. In the example of a bulk gravity theory with 3 boundaries, when the boundary vacua are specified, there is no cancellation and all the bulk geometry should be summed over. So, the partition function can not be factorized. When all the boundaries are chosen to be in the same vacuum, there is also no cancellation.}\label{vacuum}
\end{figure*}

\textbf{The classical limit where the wormholes can be seen.} The mechanism that restores the factorization is different from the way that tries to solve the puzzle by completely killing the bulk topological modes. Moreover, we can say that the bulk topological expansion, including all the connecting geometries, can be regarded as some kind of ``classical limit'' when a specific set of boundary vacua is given. As shown in Fig. \ref{vacuum}, when specific boundary vacua are given there is no cancellation, thus we have the factorization puzzle. When we ignore the possible vacuum degeneracy and calculate the partition function with the same boundary vacuum denoted by a red circle i.e. 
\be
Z\left[ {\color{red} \bigcirc} \cup  {\color{red} \bigcirc }\cup  {\color{red} \bigcirc}\right]\,,
\ee
one would never get a factorized answer. This can be represented as
\be
Z\left[ {\color{red} \bigcirc} \cup  {\color{red} \bigcirc }\cup  {\color{red} \bigcirc}\right]\neq Z\left[ {\color{red} \bigcirc}\right] \times Z\left[{\color{red} \bigcirc }\right]\times Z\left[  {\color{red} \bigcirc}\right]\,.
\ee
The above demonstration explains why we have a factorization puzzle and how we can understand the puzzle by admitting degenerate vacua. 
In the SYK model, specifying boundary vacua is equivalent to choosing a single point on the complex plane in Fig. \ref{q4}. A single point of $G_{LR}$ itself is non-zero, which is the origin of the non-factorization property in the dual theory. As argued in the previous section, every point on the complex plane comes with a phase, and the sum of those saddle points in the path integral gives zero. Choosing a specific point on the plane means that we are gauging the $\ZZ_q$ symmetry of the theory.
%So as a summary, when a specific set of boundary vacua is chosen in the JT gravity (or one single saddle in the SYK model is designated), all the bulk connecting geometries (or interactions between different replicas) can be seen. 
 %This explains what ingredient is missed in the theory with the factorization puzzle, and how to solve the puzzle by extending the Hilbert space.
~\\

\textbf{Connections between the two examples.} Are the two examples discussed in this paper, especially the symmetries $\ZZ_k$ and $\ZZ_q$, related or not? Certainly, the two theories, i.e. the JT gravity and SYK model, are closely related. Also, the cancellations in the two examples share lots of similarities. 
For instance, the wormholes are associated with extra complex phases in both cases, which directly result in the cancellation. 
%By comparing the distributions of saddles shown in Fig. \ref{q4} and the second figure of Fig. \ref{SSS}, it is quite natural to propose that symmetry $\ZZ_k$ that results in the cancellation between gravitational wormholes is closely related to the $\ZZ_q$ between $G_{LR}$ in the SYK model. 
However, despite the similarity, we have to admit that there are important differences between the two examples. In the JT gravity, the $\ZZ_k$ symmetry is merely a discrete version of a U(1) symmetry. 
In addition to the finite $k$ cases that have been shown in the paper, the cancellation is supposed to work in the large $k$ limit in the JT gravity.
While in the SYK model, $q$ is the number of interacting fermions in the Hamiltonian. It seems no clue in finding a U(1) symmetry for those $q$s. 
Moreover, the mechanism itself works for any value of $k\geq 2$ in JT gravity, but it does not work for specific values of $q$ in relation to $N$. 
The physical meaning of $q$ on the gravity side is unclear. 
So the relation between the two examples is not completely clear, and more studies to better understand the origin and connotation of the symmetries are important for the puzzle. For example, the relation between the two sides might be further clarified through research on the so-called hyperfast scrambling, i.e. large $q$ limit \cite{Jiang:2019pam,Choi:2019bmd,Gu:2021xaj,Bhattacharjee:2022ave}.

%$\mathbb{Z}_q$ in the SYK model is concrete, and the SYK cancellation happens between $q$ wormholes with different phases as can be seen from Eq. \eqref{solgs0} and Fig. \ref{q4}. 
%$q$ is the number of fermions in the Hamiltonian of the SYK model or the number of the indices of the coupling $J_{i_1\cdots i_q}$. 
%The relation between $\mathbb{Z}_k$ and $\mathbb{Z}_q$ is well-motivated because the symmetries between different saddles on both sides should be the same due to the duality between the JT gravity and the SYK model. 

\textbf{Wormhole saddles and topological changes in higher dimensions.}
The factorization puzzle is a problem in the scenario of holography, where we have the large $1/G_N$ limit in the bulk gravity theory.
In general, to understand quantum gravity problems using path integral, one needs to integrate over all the possible geometries that respect the given boundary condition. The integration is weighted by the exponential of the action. 
In the $1/G_N\to \infty$ limit, the off-shell configurations are largely suppressed, and only saddles make the dominant contributions in the path integral.
So, in this paper, we only focus on the on-shell wormhole saddles and are trying to figure out the cancellation between different saddles. 

Topological changes in higher dimensional gravity are acute \cite{Hebecker:2018ofv}, and one might need to come up with different ways to understand those geometries \cite{Betzios:2017krj,Betzios:2022oef,DeFalco:2020afv,DeFalco:2021klh,DeFalco:2021ksd}. It seems more comfortable to consider wormhole geometries in low dimensions. But one has to admit that the real problems are in higher dimensions. 
The vacuum degeneracy proposed in this paper corresponds to the twist between trumpets in JT gravity. Those extra parameters in low dimensions have no contrast with the swampland conjecture \cite{Brennan:2017rbf,Harlow:2018jwu,McNamara:2020uza}. In higher dimensions, we tend to believe that the degeneracy corresponds to some approximate global symmetries \cite{Fichet:2019ugl,Cheng:2022xyr,Cheng:2022xgm}, whose charges are soft. It was argued that those approximate ones might survive from the swampland and might provide the wormhole cancellation mechanism in higher dimensions.

\textbf{More to learn.}
The mechanism proposed here can also be regarded as a demonstration of exploring structures of quantum gravity theory from ``clouds'', just like ``Nineteenth-century clouds'' of Lord Kelvin.
For local boundary theories living in disconnected regions, there should not be interactions between these different components.
However, the presence of bulk wormholes is in contrast with the general expectation that the boundary partition function factorizes in AdS/CFT.
The above factorization puzzle is an apparent contradiction to us only because of our ignorance of the whole theory.
Here we stand with the general expectation and show that there can be interesting physics hidden in degenerate vacua structures, and when those structures are considered the puzzle can be saved. 
Moreover, we can see the puzzle because we have a traditional unique vacuum brain.

We are demonstrating two examples where the factorization can be realized in this paper. There is much more to be understood related to the quantum theory of gravity, and the real solution to the puzzle might come from different physics. 
For example, there can well be a large conspiracy related to the finite $N$ effects, and the factorization puzzle may be solved by comprehending the finite $N$ physics. Moreover, there may be a refined way of defining Euclidean path integral such that the puzzle can be understood. 
After all, physical motivation and exact exhibition of the factorization are the keys for any proposal to make sense.

%\paragraph{Acknowledgements} 
~\\~\\
\noindent\textbf{\large Acknowledgments} \\
We would like to thank Yang An, Jan de Boer, Diego Hofman, Arnab Kundu, Cheng Peng, Antonio Rotundo, and Jun-Bao Wu for their inspirational discussions.
We would also like to thank an anonymous referee for many remarks that improved the paper.
This work is supported in part by the National Natural Science Foundation of China (NSFC) under Grant No. 11905156 and No. 11935009. P.C. is also supported in part by NSFC Grant No. 12247101.
%%%%%%%%%%%%%%%%%%%%%%%%%%%%%%%%%%%%%%%%%%%%%%%%%%%%%%%%%%%%%%%%%%%%%%%%%%%%%%%%%%%%%%%%%%%%%%%%%%%%

\appendix

\section{Vacuum degeneracy of wormholes}
\label{AA}

We are going to review the path integral derivation of thermodynamics both for black holes and wormholes, following \cite{An:2023dmo}. By comparing partition functions of black holes and wormholes, we would see the vacuum degeneracy discussed in section \ref{bulk-1}.

It is well-known that the partition function of a system can be derived from the Euclidean path integral on the corresponding Euclidean manifold. 
So we usually use the Euclidean black hole geometry which is a cigar to represent the black hole partition function.
Similarly, we can use a tube, as shown in Eq. \eqref{vac}, to represent the partition function of a wormhole.
We will compare Euclidean black holes and wormholes and study the difference between them in this appendix. 

First of all, we need some basics of thermal field theory. 
Any Euclidean path integral from $\phi_1$ to $\phi_2$ can be illustrated diagrammatically as 
%\begin{figure}[H]
\be
\brat{\phi_2}{\phi_1}=\int_{\phi(0)=\phi_1}^{\phi(\tau)=\phi_2}\mD \phi~ e^{-S_E} =
\begin{matrix}
 \includegraphics[width=2.5cm]{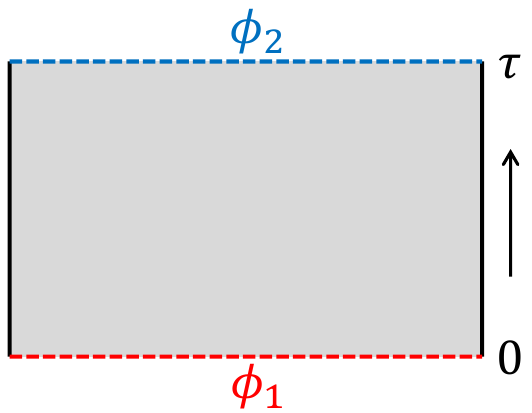}
\end{matrix}\,.\vspace{-0.5em}
\ee
%\end{figure}\vspace{-1em}
%The path integral is over the grey region, and $\phi_1$ and $\phi_2$ should serve as boundary conditions.
Moreover, the ground state $\ket{0}$ can be obtained by evolving any state $\ket{X}$ for $\infty$ amount of Euclidean time $\tau$, since the high energy eigenstates with non-zero $E_n$ are largely suppressed by $e^{-\tau E_n}$, i.e.
\be\label{infinite}
\lim_{\tau\to \infty}e^{-\tau H}\ket{X}=\ket{0}+\lim_{\tau\to \infty}\sum_{n\neq 0} e^{-\tau E_n} \ket{n}\to \ket{0}\,.
\ee
Diagrammatically, we can depict the vacuum wave functional as
%\begin{figure}[H]
\be%\vspace{-1em}
\brat{\phi_2}{0}=\int_{\phi(-\infty)}^{\phi(0)=\phi_2}\mD \phi~ e^{-S_E} =
\begin{matrix}
 \includegraphics[width=2.6cm]{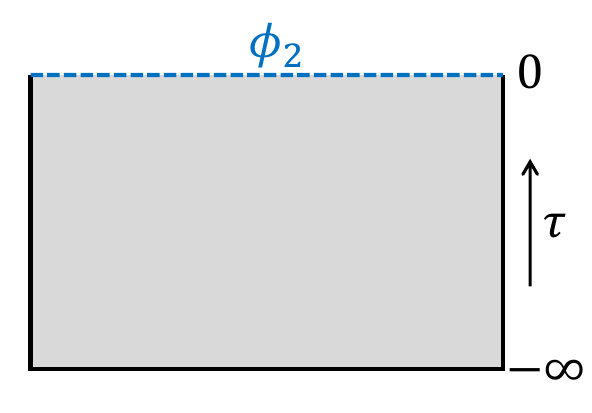}
\end{matrix}\,.\label{vacuumA}\vspace{-0.5em}
\ee
%\end{figure}
Evolving back from $\tau=0$ to $\tau=-\infty$, only the ground state can survive after the evolution, so we used the solid line in \eqref{vacuumA} to represent the vacuum state.
We will always use solid lines to denote vacua in this paper.

Now, we can consider a bipartition of space due to the presence of a horizon. We have state $\bra{\phi_2}\otimes \bra{\phi_1}$, whose vacuum wave functional measured by the overlap with vacuum state can be represented as
%\begin{figure}[H]
\be\label{TFD0}
\bra{\phi_2}\otimes \brat{\phi_1}{0}=\begin{matrix}
\includegraphics[width=2.7cm]{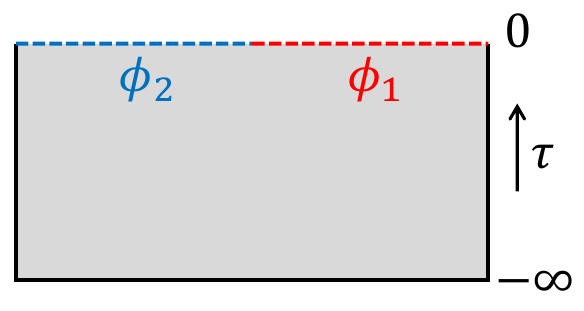}
\end{matrix}\,.
\ee
Choosing a different foliation and changing the integration variable from $\tau$ to $\theta$ shown below, we have
\be\label{TFD1}
\bra{\phi_2}\otimes \brat{\phi_1}{0}=
\begin{matrix}
\includegraphics[width=2.5cm]{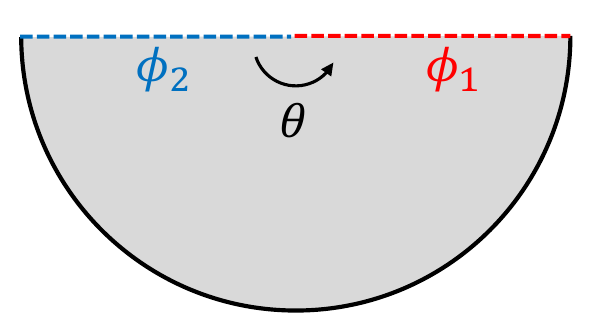}
\end{matrix}=
\bra{\phi_1}e^{-\frac{\beta}{2} H_{\theta}}\ket{\phi_2}\,,
\ee
%\end{figure}\vspace{-1em}
with $\theta \in (0,\beta/2]$.
The above expression can be further written as
\be\label{TFD2}
\bra{\phi_1}e^{-\frac{\beta}{2} H_{\theta}}\ket{\phi_2}=\sum_n e^{-\frac{\beta}{2} E_n}\brat{\phi_1}{n_1}\otimes\brat{\phi_2}{n_2}^*\,,
\ee
where $\ket{n}$ is the eigenstates of Hamiltonian $H_{\theta}$, with $H_{\theta}\ket{n}=E_n\ket{n}$ and the * symbol represents the CPT transformation.
In \eqref{TFD2}, we are evolving the wave function of $\phi_2$ to the wave function of $\phi_1$, both represented on eigenstates of $H_{\theta}$ which is $\ket{n}$. The wave function of $\phi_2$ on eigenstates $\ket{n}$ are written as $\chi_{n}[\phi_2]=\brat{\phi_2}{n_2}$, and the one for $\phi_1$ are written as $\chi_{n}[\phi_1]=\brat{\phi_1}{n_1}$. The subscripts 1 and 2 are used to denote left and right parts. 
Now, by comparing (\ref{TFD1}) and (\ref{TFD2}), the Minkowski vacuum can be expressed as the thermo-field double (TFD) state 
\be
\ket{0}=\sum_n e^{-\frac{\beta}{2}H_{\theta}}\ket{n_2}^*\otimes\ket{n_1}\equiv \ket{\text{TFD}}\,.
\ee
The reduced density matrix $\rho_R$ can be obtained by partially tracing the degrees of freedom behind the horizon in $\rho=\ketbra{\text{TFD}}$, which can be represented by a Pacman geometry
\be
\braket{\phi}{\rho_R}=%\braket{\phi}{\tr_L(\ketbra{0})}=
\begin{matrix}
 \includegraphics[width=2.1cm]{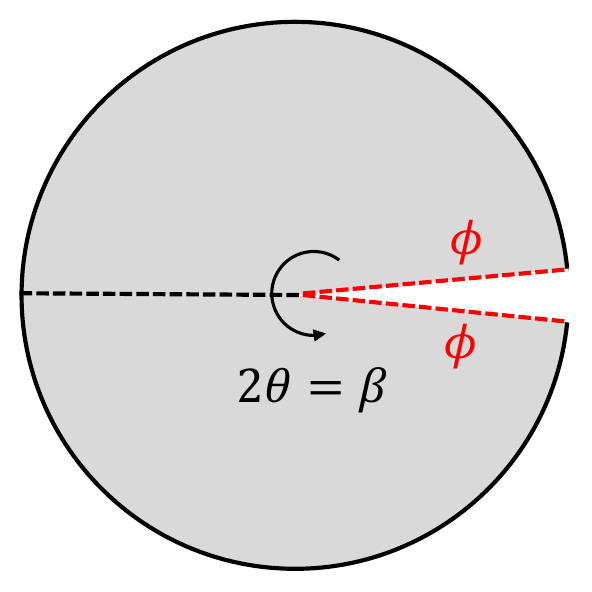}
\end{matrix}\,.\label{A8}
\ee
$\rho_R$ can be used to explain the thermal spectrum of the Unruh effect, as well as the Hawking radiation.
The partition function is a further trace of Pacman shown in \eqref{A8}, i.e. 
\be
Z=
\sum_{\phi}
\begin{matrix}
 \includegraphics[width=2.1cm]{pic/BH0}
\end{matrix}=\tr(\ketbra{\text{TFD}})\,.
\ee
For the black hole partition function, the above disk is replaced by the cigar geometry because of the curvature of the black hole. So we have 
\be
Z_{BH}=
\sum_{\phi}
\begin{matrix}
 \includegraphics[width=2.1cm]{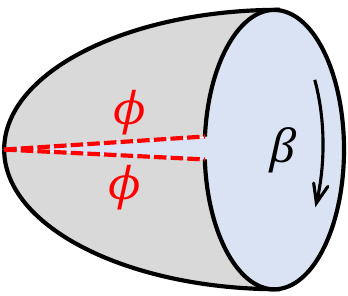}
\end{matrix}\,.\label{ZBH}
\ee

For Euclidean wormholes, TFD states are further generated to the thermo-mixed double (TMD) states \cite{Verlinde2020,Verlinde2021,Verlinde2021a}, i.e.
\be
\rho_{\text{TMD}}=\begin{matrix}
 \includegraphics[width=2.1cm]{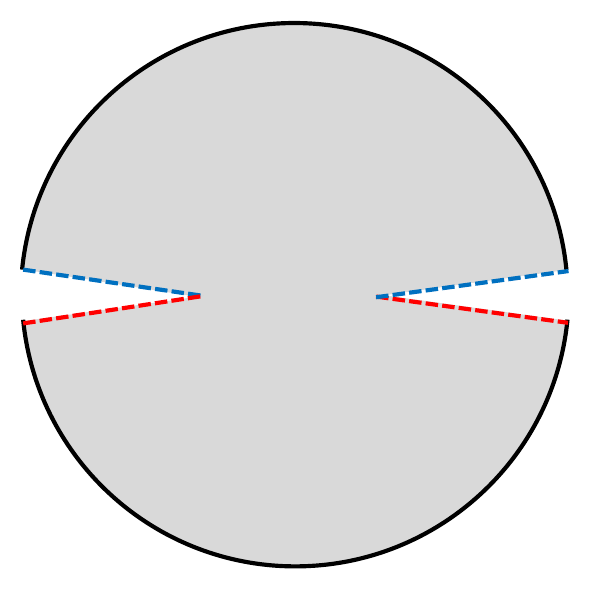}
\end{matrix}\,.
\ee
Following the same procedure in deriving the black hole thermodynamics, the wormhole partition function can be obtained by gluing two TMD states
\be
%\text{tr}[\rho^2_{\text{TMD}}]
Z_{WH} \propto \sum_{\psi_1,\psi_2}%\sum_{\phi,\bar \phi}
\begin{matrix}
\includegraphics[height=2.7cm]{pic/vac1}
\end{matrix}
=
\begin{matrix}
\includegraphics[height=2.7cm]{pic/vac2}
\end{matrix}\,.\label{vacA}\vspace{-0.5em}
\ee
The symbol $\propto$ is used due to a normalization factor.

Now, we can study the difference between the black hole and wormhole partition functions, by comparing \eqref{ZBH} and \eqref{vacA}. For black holes, the trace operation identifies vacua evolved from $\tau=-\infty$ and $\tau=\infty$. So there is no vacuum degeneracy for Euclidean black holes. However, the situation is completely different for wormholes, no trace operation identifies the vacua evolved from $\tau=-\infty$ and $\tau=\infty$. The up and down vacua $\bra{0_u}$ and $\ket{0_d}$ form their own circles. 

In 2-dimensional JT gravity, the vacuum degeneracy for wormholes was understood by the twist between the trumpets. Schematic diagrams are shown in \eqref{3WH} to illustrate the twist.
As discussed in the main text, the action $\mathcal{S}$ is used to twist $(b)$ back to $(a)$, and the cost to twist $(b)$ back is a phase related to the twisted angle. So, the phase can be regarded as the expectation value of operator $\mathcal{S}$.

%Note that the twist is characterized by a continuous U(1) parameter, while we are going to use a discrete $\ZZ_k$ symmetry to demonstrate the mechanism of realizing factorization. The reason for such a choice is that $\ZZ_k$ symmetry is enough to capture the essence of the twist shown in \eqref{3WH}, and the factorization can be related using this relatively simple version of U(1) symmetry. The $k\to \infty$ limit of $\ZZ_k$ can be regarded as the U(1) symmetry. The mechanism demonstrated in this paper should work for the $k\to \infty$ limit, so we have enough reason to believe the factorization puzzle can be understood if the symmetry is the continuous U(1) symmetry. Nevertheless, an explicit demonstration of the wormhole cancellation with continuous U(1) symmetry is worth a careful check, and we will leave it for further studies.

\section{Wormholes in SYK model with fixed coupling}
\label{BB}

We consider the following partition function in this  appendix:
\bea
&~& Z_L Z_R = \int\dd^{2N}\psi \nonumber\\&~&\times
\exp \Big{[} i^{q/2} \sum_{1<i_1<\cdots<i_q<N} J_{i_1\cdots i_q}\left(\psi^L_{i_1 \cdots i_q}+\psi^R_{i_1 \cdots i_q}\right)\Big{]}\,.\nonumber\\
\eea
It is obvious that we are allowed to insert the identity $\mathbb{I}$ in the above integral to enforce $G_{LR}=\frac{1}{N}\psi^L_i\psi^R_i$, which results in
{\small \bea
&&Z_L Z_R \nonumber\\
&&= \int\dd^{2N}\psi
~ \exp\Big{[} i^{q/2} \sum_{1<i_1<\cdots<i_q<N} J_{i_1\cdots i_q}\left(\psi^L_{i_1 \cdots i_q}+\psi^R_{i_1 \cdots i_q}\right)\Big{]}\nonumber\\ 
&& \times \int \dd G_{LR}~\delta(G_{LR}-\frac{1}{N}\psi^L_i\psi^R_i)\exp\Big{\{} \frac{N}{q}\Big{[}G_{LR}^q-(\frac{1}{N}\psi^L_i\psi^R_i)^q \Big{]}\Big{\}} \,.\nonumber\\
\eea}
One can then rewrite the delta function in its Fourier transform, with $\Sigma_{LR}$ as the Lagrangian multiplier.
Rotating the contours by $\pi/q$
\be
\Sigma_{LR}=ie^{-i\frac{\pi}{q}}\sigma_{LR}\,,~~~~~~G_{LR}=e^{i\frac{\pi}{q}}g_{LR}\,,
\ee
$Z_L Z_R$ can be written as
\be\label{ZZLR}
Z_LZ_R(J)=\int \dd \sigma_{LR}~\Psi(\sigma_{LR})\Phi(J,\sigma_{LR})\,,
\ee
with 
\be
\Psi(\sigma_{LR})=\int \frac{\dd g_{LR}}{(2\pi/N)}\exp\Big{[}N(-i\sigma g_{LR}-\frac{1}{q}g_{LR}^q)\Big{]}\,,
\ee
and 
\bea
&&\Phi(J,\sigma_{LR})=\int\dd^{2N}\psi ~\exp\Big{[}ie^{-i\pi/q}\sigma_{LR}\psi^{L}_i\psi^{R}_i\nonumber\\
&&+i^{q/2}J_{i_1\cdots i_q}(\psi^{L}_{i_1\cdots i_q}+\psi^{R}_{i_1\cdots i_q})+\frac{N}{q}(\frac{1}{N}\psi^{L}_i\psi^{R}_i)^q \Big{]}\,.
\eea
The second term $\Phi(J,\sigma_{LR})$, which depends on the coupling constants $J_{i_1\cdots i_q}$, is the part we are interested in. 
Averaging $\Phi(J,\sigma_{LR})$ over couplings, we would get 
\be\label{PhiJ}
\bracket{\Phi}_{J}=\int\dd^{2N}\psi \exp\Big{[}ie^{-i\pi/q}\sigma_{LR}\psi^{L}_i\psi^{R}_i \Big{]}=(ie^{-i\pi/q}\sigma_{LR})^N
\ee
Substituting $\bracket{\Phi}$ and $\bracket{\Psi}$ into the original partition function $Z_L Z_R$ in \eqref{ZZLR}, we get the averaged partition function  
\bea\label{Zlrj}
\bracket{Z_L Z_R}_J
&=& \frac{N}{2\pi}\int\dd g_{LR}\int\dd\sigma_{LR}\\
&\times&\exp[ N(\log(ie^{-\frac{i\pi}{q}}\sigma_{LR})-i\sigma_{LR}g_{LR}-\frac{1}{q}g_{LR}^q) ]\,.\nonumber
\eea
In the large $N$ limit, the main contributions are from the saddle points, which can be easily calculated by solving the saddle point equations. The saddle points are the $q$ solutions of
\be\label{saddleE}
(G_{LR})^q=1\,.
\ee
The $q$ solutions can be written as
\bea\label{solgs}
g^{m}_{LR} &=& e^{-i\pi/q}e^{-\frac{2\pi m i}{q}}\,,\\
\sigma^{m}_{LR} &=& -i e^{i\pi/q}e^{\frac{2\pi m i}{q}}\,,
\eea
where the $q$ solutions are labeled by $0\leq m\leq q-1$.

Now, we have got $q$ wormhole saddles in the one-time-point SYK model. However one should keep in mind that those wormhole saddles are in the averaged theory. Since the bulk-connected wormholes are the main obstacle for the factorization puzzle, we are interested in if those wormhole saddles are self-averaging or not. If the wormholes are self-averaging, we would also those wormholes in the non-averaged theories.

To see whether the wormholes are self-averaging or not, the strategy is to compare $\bracket{\Phi^2}_J$ and $\bracket{\Phi}_J^2$ around the wormhole saddles, as indicated by \cite{Saad2021}. It is easy to directly evaluate $\bracket{\Phi}_J^2$, as shown in \eqref{PhiJ}, and we have
\be
\bracket{\Phi}_{J}^2=(ie^{-i\pi/q}\sigma_{LR})^{2N}\,.
\ee
For $\bracket{\Phi^2}_J$ one need further replicate the $L$ and $R$ boundaries to four replicas $L$, $L'$, $R$ and $R'$. 
$\bracket{\Phi^2}_J$ can be denoted as
\bea
\bracket{\Phi^2}_J&=&\int\frac{\dd^4 \sigma_{AB}\dd^4 g_{AB}}{(2\pi/N)^4} \exp\Big{\{}N\Big{[}\log\big{[}(ie^{-i\pi/q} \sigma_{LR})^2\nonumber \\&~&
+X[\sigma_{AB}]\big{]}+Y[\sigma_{AB},g_{AB}]\Big{]}\Big{\}}
\eea
where $\sigma_{AB}$ and $g_{AB}$ are used to denote other collective fields rather and $\sigma_{LR}$ and $g_{LR}$. It was shown by \cite{Saad2021} that, near the wormhole saddles, the difference between $\bracket{\Phi^2}_J$ and $\bracket{\Phi}_J^2$ are small in large $N$ limit, i.e.
\be
\bracket{\Phi^2}_J\Big{|}_{\text{wormholes}}= \bracket{\Phi}_J^2\Big{|}_{\text{wormholes}}+\mathcal{O}(\frac{1}{N^{q-2}})\,.
\ee
This means that near the wormhole saddles are self-averaging in the large $N$ limit.
The conclusion can be further generated to the partition function \eqref{ZZLR}, using the self-averaging property of wormhole saddles. Near the wormhole saddles and in the large $N$ limit, the non-averaged partition function $Z_LZ_R$ can be evaluated using the averaged theory shown in \eqref{Zlrj}. Or one can say that in large $N$
\be
Z_LZ_R\Big{|}_{\text{wormholes}}=\bracket{Z_LZ_R}_J\Big{|}_{\text{wormholes}}\,.
\ee
Then, the partition function $Z_LZ_R$ can be further written as
\bea
Z_L &Z_R&\Big{|}_{\text{wormholes}}
\nonumber\\
&\approx & \frac{N}{2\pi} \sum_{m=0}^{q-1} \Big{[}(ie^{-\frac{i\pi}{q}}\sigma_{LR})^N \nonumber\\
&\times&\exp[N(-\sigma_{LR}g_{LR}-\frac{1}{q}g_{LR}^q) ]\Big{|}_{g_{LR}=g^{m}_{LR},\sigma_{LR}=\sigma^{m}_{LR} }\Big{]}\,\nonumber\\
&\propto &
\sum_{m=0}^{q-1} e^{\frac{2\pi m i}{q}N}e^{-N(1-\frac{1}{q})}\,.
\eea

%\bibliography{mybib.bib}{}
%\bibliographystyle{utphys}
\providecommand{\href}[2]{#2}\begingroup\raggedright\endgroup

\end{document}